%% file: main.tex
\documentclass[conference]{IEEEtran}
\input{pkgs}

\input{cmds}

\newcommand{\sys}{\mbox{\textsc{SpecBox}}\xspace}

\soulregister\cite7
\soulregister\autoref7
\soulregister\sys7

\begin{document}
\input{hdr}

\input{abstract}
\input{intro}
\input{bkg}
\input{threat}
\input{design}
\input{analysis}

\input{eval}

\input{discuss}

\input{relwk}
\input{conclusion}
\input{ack}

\bibliographystyle{IEEEtranS}
\bibliography{refs}

\end{document}

%% file: pkgs.tex
\usepackage{cite}
\usepackage{amsmath,amssymb,amsfonts}
\usepackage{algorithmic}
\usepackage{graphicx}
\usepackage{textcomp}
\usepackage[hyphens]{url}
\usepackage{listings}
\usepackage{float}
\usepackage{pifont}

\usepackage[hyphens]{url}
\usepackage[breaklinks,colorlinks]{hyperref}
\usepackage[usenames,dvipsnames]{xcolor}
\hypersetup{citecolor=blue,linkcolor=blue}
\usepackage{amsmath,amsopn,amssymb}
\usepackage{subfigure}
\usepackage{endnotes,microtype,xspace,graphicx,fancyvrb,multirow,diagbox}
\usepackage{booktabs}
\usepackage[T1]{fontenc}
\usepackage{fancyhdr,lastpage}
\usepackage{enumitem}
\usepackage[labelfont=bf,font=small,skip=5pt]{caption}
\usepackage{pifont}

\usepackage{fp}
\usepackage{siunitx}

\usepackage{balance}

\usepackage{soul}

\def\BibTeX{{\rm B\kern-.05em{\sc i\kern-.025em b}\kern-.08em T\kern-.1667em\lower.7ex\hbox{E}\kern-.125emX}}

\fancypagestyle{firstpage}{
  \fancyhf{}

  \fancyhead[C]{\normalsize{ISCA 2021 Submission
      \textbf{\#394} \\ Confidential Draft: DO NOT DISTRIBUTE}} 
  \fancyfoot[C]{\thepage}
}

%% file: cmds.tex
\definecolor{mygreen}{rgb}{0,0.6,0}
\definecolor{mygray}{rgb}{0.5,0.5,0.5}
\definecolor{mymauve}{rgb}{0.58,0,0.82}
\lstdefinelanguage{myC}{
	backgroundcolor=\color{white},      
	basicstyle=\footnotesize\ttfamily,
	columns=fullflexible,
	tabsize=4,
	breaklines=true,               
	captionpos=b,                  
	commentstyle=\color{mygreen},  
	escapeinside={\%*}{*)},        
	keywordstyle=\color{blue},     
	morekeywords={foreach, int, char, ulong, return, asm, if, else, in, delete, ADDR, ADDR_SET, void, update, with, then, unsigned, long, define, uint8},
	stringstyle=\color{mymauve}\ttfamily,  
	frame=single,
	rulesepcolor=\color{red!20!green!20!blue!20},
	numbersep=5pt, 
	numberstyle=\tiny,
	numbers=left, 
	morecomment=[l]//,%
	morecomment=[s]{/*}{*/},%
	morestring=[b]",%
	morestring=[b]',%
	xleftmargin = 1em,
}

\newenvironment{packeditemize}{
\begin{list}{$\bullet$}{
\setlength{\labelwidth}{8pt}
\setlength{\itemsep}{0pt}
\setlength{\leftmargin}{\labelwidth}
\addtolength{\leftmargin}{\labelsep}
\setlength{\parindent}{0pt}
\setlength{\listparindent}{\parindent}
\setlength{\parsep}{2pt}
\setlength{\topsep}{3pt}}}{\end{list}}

\newcommand{\Autoref}[1]{%
  \begingroup%
  \def\chapterautorefname{Chapter}%
  \def\sectionautorefname{Section}%
  \def\subsectionautorefname{Subsection}%
  \autoref{#1}%
  \endgroup%
}
\newcommand{\whiteding}[1]{\ding{\numexpr191+#1\relax}}

\newcommand{\bheading}[1]{{\vspace{4pt}\noindent{\textbf{#1}}}}
\newcommand{\tabincell}[2]{\begin{tabular}{@{}#1@{}}#2\end{tabular}}
\newcommand*\rot[1]{\rotatebox{90}{#1}}

%% file: hdr.tex
\title{\sys: A Label-Based Transparent Speculation Scheme Against Transient Execution Attacks} 

\author{
    \IEEEauthorblockN{Bowen Tang$^{1,2}$ , Chenggang Wu$^{1,2}$, Zhe Wang$^{1,2}$\thanks{* Corresponding Author: Zhe Wang}, Lichen Jia$^{1,2}$, Pen-Chung Yew$^{3}$, Yueqiang Cheng$^{4}$, \\ Yinqian Zhang$^{5}$, Chenxi Wang$^{6}$, Guoqing Harry Xu$^{6}$}
    
    \IEEEauthorblockA{$^1$State Key Laboratory of Computer Architecture, Institute of Computing Technology, Chinese Academy of Sciences, \\ $^2$University of the Chinese Academy of Sciences, $^3$University of Minnesota at Twin Cities, $^4$NIO Security Research, \\ $^5$Southern University of Science and Technology, $^6$University of California, Los Angeles}
    
    \IEEEauthorblockA{$^{1}$\{tangbowen, wucg, wangzhe12, jialichen\}@ict.ac.cn, $^{3}$yew@emn.edu, $^{4}$strongerwill@gmail.com, \\ $^{5}$yinqianz@acm.org, $^{6}$\{wangchenxi, harryxu\}@cs.ucla.edu}
}

\maketitle
\pagestyle{plain}

%% file: abstract.tex
\begin{abstract}

Speculative execution techniques have been a cornerstone of modern processors to improve instruction-level parallelism. However, recent studies showed that this kind of techniques could be exploited by attackers to leak secret data via transient execution attacks, such as Spectre. 
Many defenses are proposed to address this problem, but they all face various challenges: (1) Tracking data flow in the instruction pipeline could comprehensively address this problem, but it could cause pipeline stalls and incur high performance overhead; (2) Making side effect of speculative execution imperceptible to attackers, but it often needs additional storage components and complicated data movement operations. 
In this paper, we propose a \emph{label-based transparent speculation} scheme called \sys. It dynamically partitions the cache system to isolate speculative data and non-speculative data, which can prevent transient execution from being observed by subsequent execution. Moreover, it uses thread ownership semaphores to prevent speculative data from being accessed across cores. In addition, \sys also enhances the auxiliary components in the cache system against transient execution attacks, such as hardware prefetcher. Our security analysis shows that \sys is secure and the performance evaluation shows that the performance overhead on SPEC CPU 2006 and PARSEC-3.0 benchmarks is small.

\end{abstract}

%% file: intro.tex
\section{Introduction}\label{sec:intro}
Five decades of exponential growth in processor performance has led to today's ubiquitous computation infrastructure. At the heart of such rapid growth are many optimizations employed in today's processors. Among them, using speculative execution to alleviate pipeline stalls caused by control flow transfer and memory access is one of the most effective optimizations in modern processors. However, recent studies have shown that this kind of techniques can introduce security vulnerabilities and be exploited by attackers via \emph{transient execution attacks (TEAs)} such as the well-known Spectre~\cite{Spectre}. These vulnerabilities are widely present in billions of processors produced by mainstream manufacturers such as Intel, AMD and ARM.

\autoref{list:poc} is a proof-of-concept (PoC) code in the Spectre attack. First, the attacker steers the index value to be less than \texttt{SIZE}, thereby training the branch predictor to choose the fall-through branch in Line 5. After that, the attacker steers the index value to be greater than \texttt{SIZE}, which leads to an out-of-bound access to the secret during the speculative execution. Then, the attacker utilizes the secret to index the \texttt{dummy} array. This will load the \texttt{dummy[val*64]} item into the cache. Eventually, when the processor determines that it is a mis-prediction in Line 5, it will squash the instructions of Lines 6-7 and follow the correct path. In the current design, the processor will not clean up the side effects in the cache (i.e. the data brought in during the mis-speculative execution). The attacker can scan the \texttt{dummy} array in the cache, and measure the access time of each array element. According to the access latency, the attacker can decide which item has been loaded into the cache, and thereby infer the secret value. 

\begin{lstlisting}[language=myC,caption=A proof-of-concept (PoC) code of the Spectre attack.,label=list:poc]
#define SIZE 10
uint8 array[SIZE];
uint8 dummy[256*64];
uint8 victim(int index) {
	if (index < SIZE) {
		uint8 val = array[index];
		return dummy[val*64];
	} else return 0;
}
victim(8);// Step1: train the branch
flush_cache();// Step2: prepare cache layout
victim(100); // Step3: access the secret
timing_dummy(); // Step4: measure cache layout
\end{lstlisting}

To mitigate TEAs, several approaches have been proposed in recent years. The first category of approaches is to delay the use of the data until the instructions that produce the data are no longer speculative, i.e. the data is no longer ``tainted'' \cite{STT} and ``safe'' to use. For the example in \autoref{list:poc}, the out-of-bound index value in Line 6 cannot be used to access the \texttt{dummy} array until the branch condition in Line 5 is resolved. The representative schemes are STT~\cite{STT}, SDO~\cite{SDO}, NDA~\cite{NDA}, ConditionalSpec~\cite{ConditionalSpec}, Delay-on-Miss~\cite{delayonmiss} and SpecShield~\cite{SpecShield}. These defenses comprehensively prevent the execution of instructions that may cause information leackage. However, delaying data propagation can cause pipeline stalls, which may incur high performance overhead~\cite{STT, NDA, delayonmiss, SpecShield}.


The second category is to keep speculative execution non-blocking but make it "invisible" to the subsequent instructions if it fails. In \autoref{list:poc}, the array elements of \texttt{dummy} loaded into the cache during the speculative execution will be removed and cannot be accessed in the \texttt{timing\_dummy()}. From the perspective of microarchitecture designers, this approach is preferable because of its compatibility with other optimization mechanisms that are critical to processor performance~\cite{Muontrap}. The representative works are SafeSpec~\cite{SafeSpec}, InvisiSpec~\cite{InvisiSpec}, CleanupSpec~\cite{CleanupSpec} and MuonTrap~\cite{Muontrap}. 

To achieve the invisible speculative execution, InvisiSpec, SafeSpec, and Munontrap choose to add an extra storage to keep the speculatively installed data. When the speculative instructions are committed, the speculatively installed data will be \emph{re-installed} into the cache hierarchy from the memory system or the newly added storage to make it visible. In contrast, CleanupSpec allows the data to be installed in the cache hierarchy during the speculative execution, and the replaced data is stored into a newly added storage. The replaced data will be \emph{re-installed} into the cache hierarchy to rollback the cache state only when speculation fails. Since most speculation will succeed, CleanupSpec has a higher performance in general~\cite{CleanupSpec}. But, the \emph{re-install} operations are required no matter whether the speculation succeeds or fails. Such additional data movement on the cache hierarchy can reduce its benefit and degrade its performance. 

In this paper, we re-exam the invisible speculative execution strategy along with the cache design, and propose a new transparent speculation scheme, called \sys. It modifies the cache to support the invisible speculative access efficiently. The speculative data and the non-speculative data are distinguished in the cache, so the extra storage and data movement are no longer needed. To achieve this, \sys divides each cache set into two domains. Each cache line in the set is distinguished by a 1-bit label to indicate which domain it is in. The \emph{temporary domain} contains the speculative data and the \emph{persistent domain} contains the non-speculative data. When the speculation fails, the speculatively installed cache lines in the temporary domain will be invalidated. When the speculation succeeds, \sys only needs to flip the bits to switch the corresponding cache lines from the temporary domain to the persistent domain. Thus, it totally avoids the data movement required in other schemes, and hence, improves the performance.

In addition to isolating the speculative data in the single-core, we also extend the \emph{label-based method} to the multi-core environment and make the speculative data invisible across cores. \emph{Physically} isolating hardware threads' speculative data from each other can also achieve this, but it limits the scalability and has low resource utilization. The core idea in \sys is \emph{to dynamically mark the thread ownership of each shared cache line in the temporary domain, and emulate a behavior that is similar to accessing the thread-private cache}. To achieve that, \sys attaches each cache line with an N-bit label, with each bit bound to a hardware thread (HT). If the bit corresponding to a HT is set, it means this HT owns this cache line. When a HT accesses a cache line which is not owned by this HT in the temporary domain, \sys will simulate a latency equivalent to a miss and then sets the corresponding bit, even if it has been speculatively installed by other threads. When a HT evicts a cache line owned by other HTs in the temporary domain, \sys will just reset its corresponding bit. The set/reset actions are similar to the P/V semaphore operations in parallel programming, so we call the N-bit label \textit{thread-owner semaphore (TOS)}.

To summarize, we make the following contributions:

\begin{enumerate}
\item We propose a new \emph{label-based transparent speculation} scheme, called \sys, against transient execution attacks.
It leverages a cache partitioning approach on speculative data, which eliminates the need for data movement during the switch between speculative and non-speculative execution.

\item We propose a thread-ownership semaphore to logically isolate shared data among threads to prevent side channels to be formed in the shared cache.

\item The security analysis shows that the proposed \emph{label-based transparent speculation} scheme is secure. And the performance evaluation shows that the overhead of \sys is substantially lower than that of STT, InvisiSpec and CleanupSpec on SPEC CPU 2006 and PARSEC-3.0 benchmarks.
\end{enumerate}


%% file: bkg.tex
\section{Background}\label{sec:bkg}

\subsection{Speculative Execution}
Speculative execution techniques, such as branch prediction~\cite{BTB} and memory disambiguation~\cite{memory-disambiguation}, are commonly used on modern out-of-order processors to improve performance. To avoid pipeline stalls, the processor continues to speculatively execute instructions beyond a branch instruction along a predicted path before the branch condition is resolved. If the prediction fails, the mis-speculated instructions will be squashed. A \emph{re-order buffer} (ROB) is used to maintain correctness after the out-of-order execution. When an instruction reaches the head of the ROB and has completed its execution, it updates the machine state and releases its held resources, this process is called \emph{commit}. An instruction not yet committed is called an \emph{in-flight instruction}. 

On modern processors, the side effects caused by the mis-speculated instructions such as the data brought into the cache memory during the speculative execution are not cleaned up. It will not affect the correctness of the program execution, but could impact the timing of subsequent instructions and can be used by attackers to create covert channels.

\subsection{Timing-Based Covert Channel}\label{sec:covert}
The core logic of a transient execution attack can be divided into three steps: a) Accessing the secret through speculative execution; b) Encoding the secret through side effects that affect the subsequent execution timing; c) Decoding the secret through executing some operations and time their execution~\cite{TEAEvaluation}. The first two steps are collectively referred to as the \emph{sender}, and the third step is referred to as the \emph{receiver}. The transmission channel between the sender and receiver is known as \emph{timing-based covert channel}~\cite{Timing_Survey}.

According to the length of the information carrier's life span, timing-based covert channels can be grouped into two types: the \emph{persistent} and the \emph{volatile}~\cite{TEA_Survey}. The information carrier in a \emph{persistent covert channel} is usually the layout of a storage unit, such as the cache~\cite{cache}, \emph{translation lookaside buffer} (TLB)~\cite{TLB}, and \emph{branch target buffer} (BTB)~\cite{BTB}, or a state change such as the on/off state of high bits in the AVX2 vector register~\cite{schwarz2019netspectre}. 
In a \emph{persistent covert channel}, the side effects encoded by the sender will exist for a relatively long period of time to allow the receiver to extract the secret value. In contrast, the information carrier of a \emph{volatile covert channel} is usually a shared resource between threads, such as floating-point unit (FPU)~\cite{lazyfp}, execution unit port~\cite{port-contention1} and memory bus~\cite{memorybus}. The side effect exploited by the sender is to delay the execution time of the receiver by competing for such resources. Because resource competition is transient, the receiver must measure the timing while the sender is accessing the share resources. In this case, the sender and the receiver must be synchronized, which increases the difficulty for the attacker.

\subsection{Transient Execution Attacks}
Transient execution attacks (TEA) can be divided into two categories: the \emph{Spectre-type} and the \emph{Meltdown-type} attacks \cite{TEAEvaluation}. Attackers can directly access unauthorized memory locations or registers during the out-of-order execution in a Meltdown-type attack. Unlike the Spectre-type attacks, which primarily exploiting speculative execution, the Meltdown-type vulnerabilities are mostly due to unintended hardware bugs. They can thus be fixed in the hardware. For Spectre-type attacks, the attacker will first train the prediction units, such as the \emph{pattern history table} (PHT)~\cite{PHT}, BTB~\cite{BTB} and \emph{return stack buffer} (RSB)~\cite{RSB}. After that, the attacker will bypass the protection code to execute a wrongly speculative path, such as a bound check~\cite{Spectre-PHT1}, data cleanup~\cite{Spectre-STL} and stack pointer switching~\cite{swapgs}, and then access the secret. Finally, the attacker will transmit the secret through a timing-based covert channel.


\subsection{Existing Defenses Against TEAs}
\label{sec:defenses}
For TEAs, a common defense is to prevent sensitive data from being transmitted to covert channels. SpecShield~\cite{SpecShield} checks whether there is an unresolved branch or an instruction that triggers an exception before a load instruction, and then determines whether the loaded data can be passed to subsequent instructions. ConditionalSpec~\cite{ConditionalSpec} puts forward the concept and a detection method of ``safe dependence'', and proposes an S-pattern filtering strategy to improve the detection efficiency according to the characteristics of TEAs. NDA~\cite{NDA} and STT~\cite{STT} use dataflow tracking similar to taint propagation, and track those instructions that may cause information leakage. Those instructions are forced to delay until their dependent instructions become safe. To reduce the overhead caused by the delay and the tracking analysis, SDO~\cite{SDO} adds a safe value prediction mechanism based on STT. But the performance overhead is still high due to the pipeline stalls that cannot be avoided.

Make speculative execution ``invisible'' by cleaning up all side effects after speculative execution could avoid the pipeline stalls, and prevent side effects from being used by attackers to transmit the secret data.
InvisiSpec~\cite{InvisiSpec} chooses to add a \emph{speculative buffer} to keep the speculatively installed data. Similarly, SafeSpec~\cite{SafeSpec} and MuonTrap~\cite{Muontrap} use a \emph{shadow cache} and a \emph{non-inclusive L0 cache}, respectively. If speculation succeeds, the speculative installed data will be re-installed into the newly introduced storage; if speculation fails, the speculatively installed data in these storages will be cleaned up. Since most speculation will succeed, these methods will introduce non-trivial overhead. Hence, CleanupSpec~\cite{CleanupSpec} was proposed to allow the data be speculatively installed into the original cache, and only rollbacks the cache state via re-installing the replaced data when speculation fails.

Actually, the \emph{re-install} operations are required no matter whether the speculation succeeds or fails. Such data movement triggered by the re-install operations will degrade performance. It motivates us to develop a scheme that can perform an "invisible" speculative execution inside the cache system without requiring any ``data movement'' during the switch between speculative and non-speculative execution.


%% file: threat.tex
\section{Threat Model}\label{sec:threat}


In this paper, we mainly focus on the cache system that is vulnerable to the transient execution attacks. Here, we assume attackers have the following abilities:

\begin{packeditemize}
\item	Ability to train the control flow prediction units and memory disambiguation units in order to exploit all Spectre- and Meltdown-like vulnerabilities.

\item	Ability to find/execute the gadgets and to access/encode secret data. The secret data here refers to various protected memory locations and registers.

\item	Ability to know the cache indexing method and its replacement strategy, which means the attacker can install or evict a cache line at any location in the cache.

\item	Ability to launch multiple threads located on a SMT core and/or different cores, and control their interleavings in order to transmit the stolen secret data via cache.
\end{packeditemize}

Our goal is to defeat TEAs that use \emph{persistent covert channels}. For TEAs using \emph{volatile covert channels}, such as port contention~\cite{SMoTherSpectre,port-contention1}, it can be mitigated by turning off SMT, using security-sensitive thread scheduling~\cite{russo2006securing}, or using time-division multiple accessing (TDMA) on shared resources~\cite{SMT-COP}. These methods are orthogonal to our approach. 

Also, we do not consider \emph{physical} covert channels, such as electromagnetic signals~\cite{electromagnetic-signal} and power consumption~\cite{power-consumption,power-consumption1}. These channels are generally noisier, requiring longer time for attackers to observe the effects. 
At present, these types of covert channels cannot be used in transient execution attacks.

%% file: design.tex
\section{Problem Analysis}\label{sec:problem}


\subsection{Encoding Methods of Cache Covert Channel} 
As mentioned in \autoref{sec:covert}, the \emph{sender} illegally accesses the secret and encodes it into the cache layout; the \emph{receiver} infers the secret via timing access to the cache. The encoding schemes can have the following two types.


\begin{packeditemize}
\item \textbf{Acceleration-based encoding} The \emph{sender} first evicts an item from the cache, then decides whether to re-install this item or not according to the speculative accessed secret value. Taken a 1-bit secret as an example, if the secret value is 1, the item will be re-installed back. Therefore, when the \emph{receiver} accesses this item again, the timing will be short due to the cache hit. The \emph{receiver} can reason about the secret value being 1. Otherwise, it is 0. 


\item \textbf{Deceleration-based encoding} The \emph{sender} first installs an item into the cache, then decides whether to replace this item with another item or not according to the secret value. Also taken a 1-bit secret as an example, if the secret value is 1, the new item will replace the original item. Therefore, when the \emph{receiver} accesses the original item again, the timing will be long due to the cache miss. The \emph{receiver} can conclude that the secret value is 1. Otherwise, it is 0. 

\end{packeditemize}

\vspace{-0.8em}

\subsection{Attack Modes}

\begin{figure}[!t]
	\centering
	\includegraphics[width=\columnwidth]{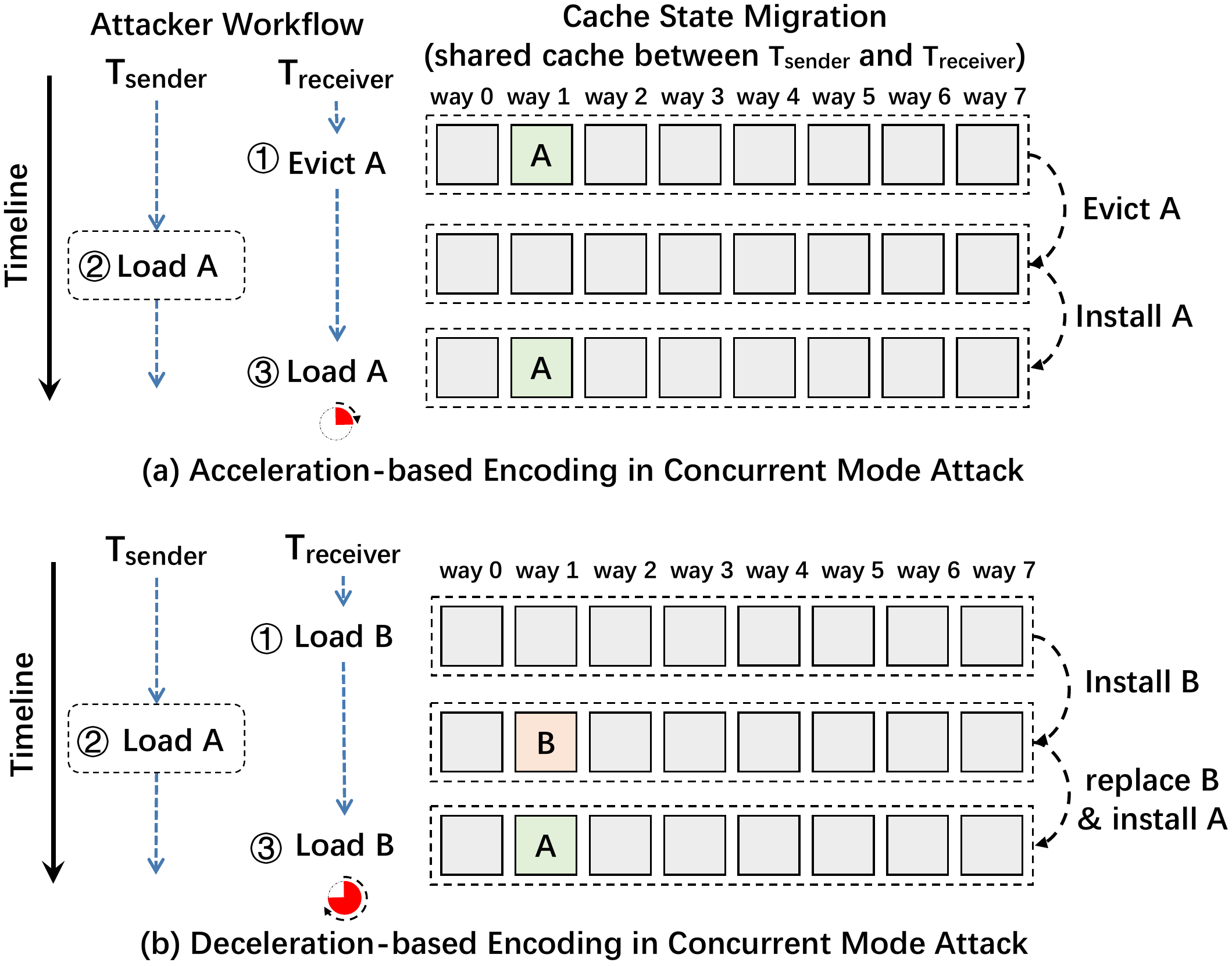}
	\caption{The Encoding Methods in Concurrent Mode Attacks.}
	\vspace{-1.8em}
	\label{fig:concurrent_mode}
\end{figure}

In most common attack scenarios, the above two encoding methods are performed in a \emph{serialized} manner, that is, the receiver’s decoding will not begin until the sender's encoding has been completed. 
However, in multi-thread applications, the encoding can be conducted in a \emph{concurrent} manner. The attacker can launch two concurrent threads located on different cores or a SMT core, with one thread acting as the sender and the other as the receiver. 
Using browser as an example, multiple threads of a malicious plug-ins can occupy different cores. Their targets can be user’s passwords or cookies, which are protected by the sandbox in the browser. The malicious plug-ins cannot access them directly, but can exploit the \emph{Spectre-like} vulnerability to access them in a wrong-path speculative execution, and leverage the shared cache as a covert channel to transmit them.

\autoref{fig:concurrent_mode} shows how the two encoding methods can be implemented in a concurrent mode. For the acceleration-based method, the $T_{receiver}$ first evicts the item and then the $T_{sender}$ speculatively accesses the secret and re-installs the item according to the 1-bit secret value (assuming the value is 1). Immediately afterwards, the receiver access the item again and measures the access time. At that time, the speculation of $T_{sender}$ has not been committed yet, which means other defenses, such as \emph{cleanup} or \emph{rollback}, has not yet to be carried out. 
The $T_{receiver}$'s revisit will be hit. Similarly, for the deceleration-based method, after the $T_{sender}$ has replaced the item primed by the speculative install of $T_{receiver}$, and before the process of $T_{sender}$ is committed, the $T_{receiver}$ will experience a cache miss when the replaced item is accessed again.

\section{Our Solution}\label{sec:design}
To defeat the above attacks in both serial and concurrent modes and keep the speculative execution efficient, we proposed a \emph{label-based transparent speculation} scheme, called \sys. In \sys, all speculative (i.e. \emph{in-flight}) operations, such as loads/stores and instruction fetching that may affect the cache system, are regarded as unsafe until they reach the head of ROB and are committed. As mentioned earlier, the core ideas of \sys consists of two major components.

\begin{figure}[!t]
	\centering
	\includegraphics[width=0.95\columnwidth]{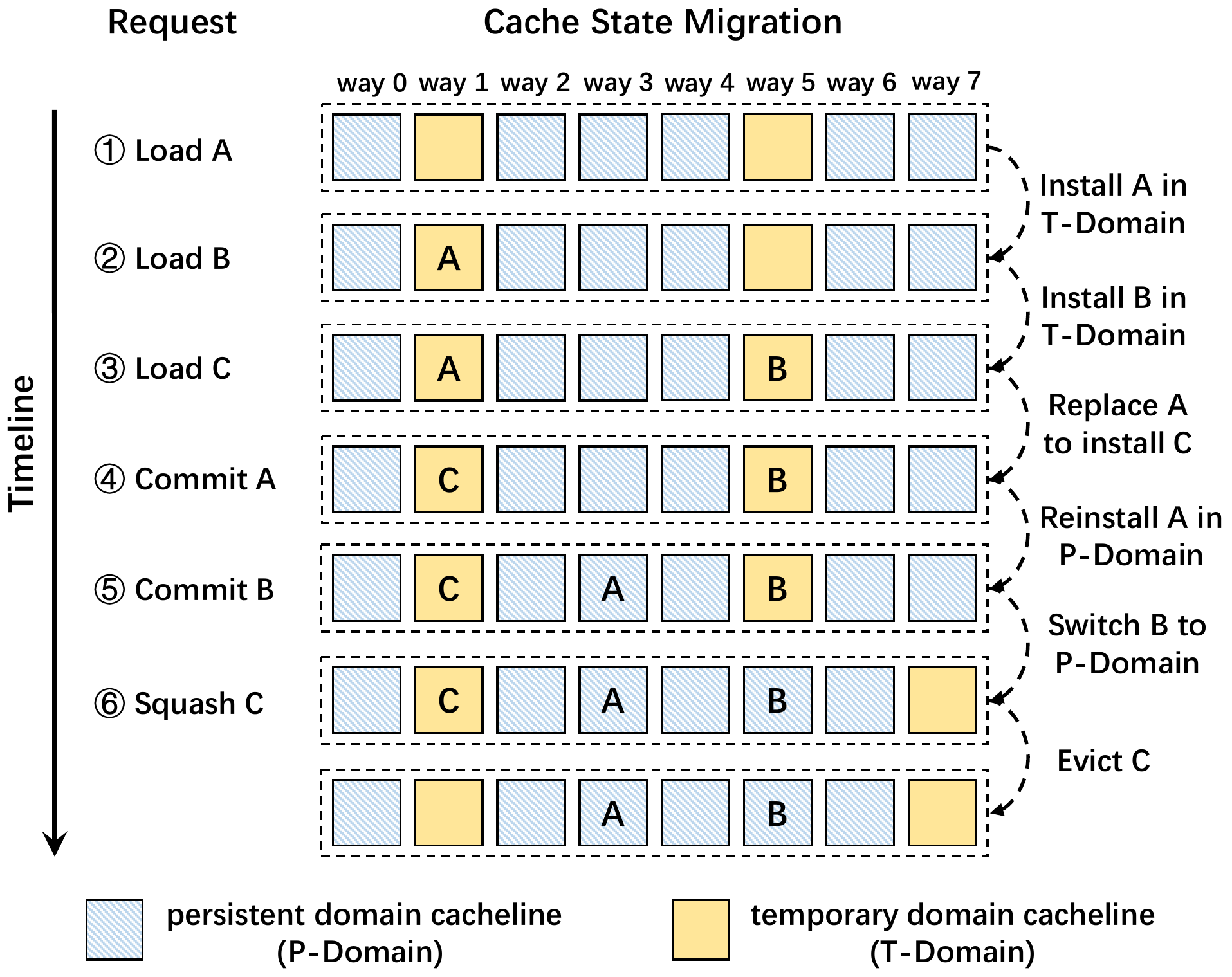}
	\caption{An example of the state migration on one cache set.}
	\vspace{-1.8em}
	\label{fig:partition}
\end{figure}

\subsection{Domain Partition}\label{sec:partition}

\subsubsection{Partitioning Scheme} 
As shown in \autoref{fig:partition}, \sys partitions each cache set into two domains: (1) the \emph{temporary domain} that contains the data installed by the \emph{in-flight} operations, and (2) the \emph{persistent domain} that contains the data installed by the \emph{committed} operations. The partition is implemented by adding a 1-bit label, called T/P (Temporary/Persistent) flag, on each cache line’s tag area. If the cache line's T/P flag is 0, it belongs to the temporary domain; otherwise, it belongs to the persistent domain. 
By modifying the T/P flag, \sys can switch the domain the cache line belongs to. The domain needs not be contiguous in physical space but need to have a pre-determined capacity, so that it will not compete against each other. When installing a new cache line or switching a cache line’s domain from the other domain, \sys will check whether the current domain has reached its capacity. If it has, it will wake up the replacement module to select a cache line within the domain and replace it.

\subsubsection{Access Control}
\begin{figure}[!t]
	\centering
	\includegraphics[width=\columnwidth]{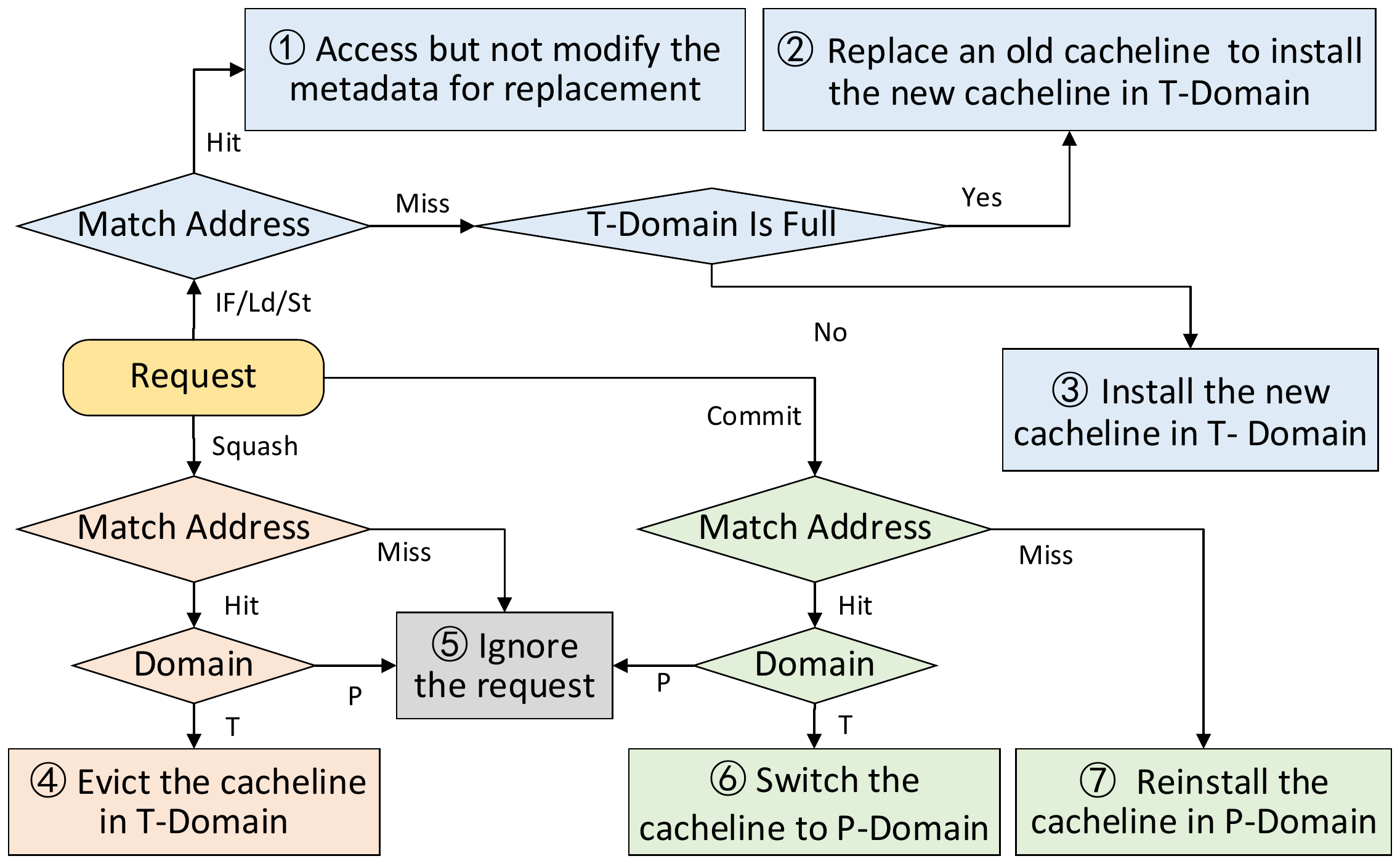}
	\caption{Data access flow in temporary/persistent domains}
	\vspace{-1.8em}
	\label{fig:data_access}
\end{figure}

\Autoref{fig:data_access} shows the access flow on a cache with the proposed partitioning scheme. 
The access flow starts from the yellow box marked as ``\emph{Request}''. \whiteding{1} If the request is an \emph{in-flight} operation and is a cache ``hit'', regardless of whether it hits the temporary domain or the persistent domain, the operation can directly access the data without modifying the metadata for cache replacement. \whiteding{2} If it is a cache miss and the temporary domain is full, the cache will select one victim cache line from the temporary domain and replace it. \whiteding{3} If the temporary domain is not full, it will install a new cache line in the temporary domain.  
\whiteding{4} When an \emph{in-flight} access operation is squashed, the cache will receive a squash request from CPU. It checks whether the data installed by the operation is still in the temporary domain and evicts the entry if so. \whiteding{5} If the data installed is not in the temporary domain, it just ignores the request. \whiteding{6} When an \emph{in-flight} operation is committed, the cache will receive a commit request from CPU. It will check whether the data installed by the operation is in the temporary domain and if so, the cache converts the cache line to the persistent domain by setting the T/P flag. Meanwhile, to keep the capacity constant, the cache will evict one victim cache line from the persistent domain and switch it to the temporary domain. If the data installed by the operation is already in the persistent domain, it follows case \whiteding{5} and ignores the request. \whiteding{7} If the data is neither in the persistent domain nor in the temporary domain, the cache will re-install it in the persistent domain.

\Autoref{fig:partition} depicts an example of the state transition (i.e. domain switches) on one cache set during some request accesses. This set is 8-way associative, and the capacity ratio of the temporary and persistent domains is 2:6. Firstly, the cache receives two consecutive load requests, i.e., \emph{Load A} (\whiteding{1}) and \emph{Load B} (\whiteding{2}) in the figure. Since the A and B are missing in the cache, these requests will trigger cache miss handling events. After A and B are fetched from the next level cache or memory, they are installed into the temporary domain; Then, the cache receives the \emph{Load C} request (\whiteding{3}) and C is also missing. But, there is no more cache line in the temporary domain in the set, so the cache chooses to replace A with C in accordance with the replacement algorithm. After that, the \emph{in-flight} operation to which the \emph{Load A} request belongs is committed (\whiteding{4}), the cache receives a \emph{Commit A} request. But, A has been evicted from the set. The cache thus selects a victim cache line from the persistent domain and re-installs A. After that, the cache receives a \emph{Commit B} request (\whiteding{5}). It first selects a victim cache line from the persistent domain to evict, and switch it to the temporary domain, and then switch B to the persistent domain. Finally, the cache receives a \emph{Squash C} request (\whiteding{6}). It evicts the cache line that contains C. 

\subsubsection{Notifier}
\begin{figure}[!t]
	\centering
	\includegraphics[width=0.9\columnwidth]{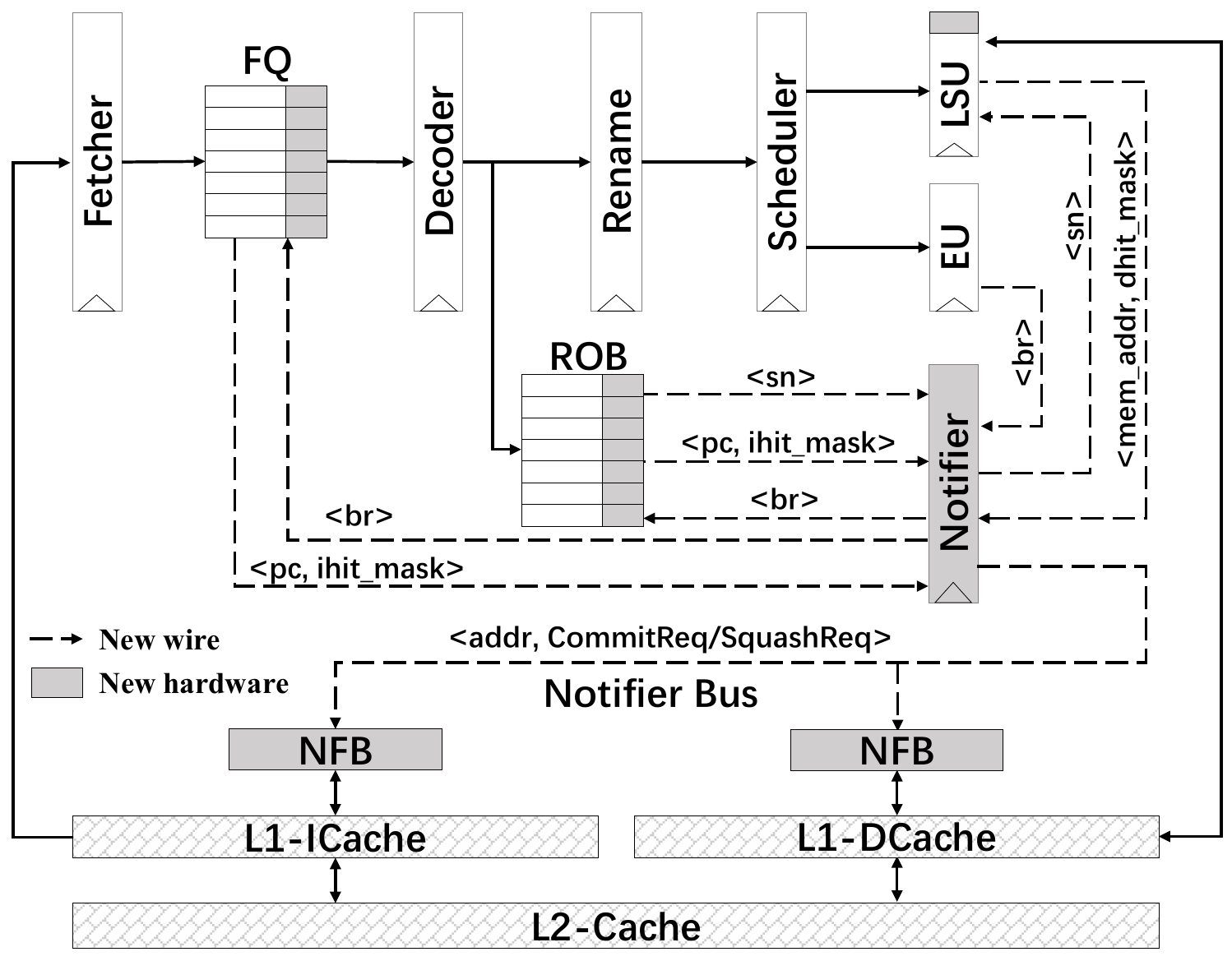}
	\caption{The enhanced pipeline and cache system with the \emph{notifier}.}
	\vspace{-1.8em}
	\label{fig:notifier}
\end{figure}

We added a notifier in the commit stage of the pipeline (shown in \autoref{fig:notifier}).
It notifies the cache system to switch domain or clean up the corresponding cache line when an \emph{in-flight} operation is committed or squashed. The notifier performs different actions according to the types of operations.

\bheading{For Load/Store.} 
In order to know which level of the cache hierarchy has been accessed by a load/store instruction, \sys adds some bits as \emph{dhit\_mask} in each entry of the \emph{Load/Store Unit} (LSU).
They record whether the access to the cache level has been hit or not. When the ROB commits or squashes a load/store instruction, it will signal the notifier with the \emph{sequence\_number} (sn) of ROB. Subsequently, the notifier acquires the LSQ with the \emph{sn} for the \emph{memory\_address} and \emph{dhit\_mask}, and then generates the notification request.

\bheading{For IFetch.} 
Although the CPU does not track the instruction fetching process directly, we could still infer the process by analyzing the effects of instruction fetching. One portion of the instructions have been decoded and dispatched to the ROB, and the other portion have not yet been decoded in the \emph{Fetch Queue} (FQ). Similar to LSU, \sys adds the \emph{ihit\_mask} bits in the entries of the FQ and ROB. When the \emph{Execution Unit} (EU) resolves a branch and accepts/rejects the prediction of the branch, it will signal the notifier with the \emph{resolved\_branch} (br) to obtain the \emph{PC\_address} and \emph{ihit\_mask} of all the fetched instructions located in that branch from the FQ and ROB, and then generate the notification requests.

In order not to disturb the instruction fetching and load/store operations, the notification requests will be first sent to the L1 I-cache and L1 D-cache through the \emph{Notifier Bus}. Then the L1 cache controllers filters these requests and forward the requests to the lower-level cache through the interconnect. Moreover, in our study, we find that there is a significant amount of notification requests for the same cache line caused by contiguous cache accesses (mostly due to instruction fetching). Therefore, \sys adds a \emph{Notification Fill Buffer} (NFB) in \emph{Notifier Bus} with 16 entries (i.e. the size of the cache line) to merge the redundant requests to the same cache line.


\subsection{Thread Ownership Semaphore (TOS)}\label{sec:tos}
Inter-thread isolation by allocating a private memory region to each thread could make the speculation invisible across cores, but it will lower the resource utilization and limit the scalability. In this paper, we propose a label-based solution called \emph{thread ownership semaphore} (TOS). It is based two insights: 1) it's not necessary to keep it invisible for the data installed by the \emph{committed} operations; 2) The data installed by the \emph{in-flight} operations of one thread need not be forever invisible to other threads. If another thread has also accessed the data during the \emph{in-flight} operations, it should be safe to share the data between the two threads. So, we conclude that the problem of maintaining invisibility between the threads is equivalent to the problem of tracking the ownership of each thread and emulate its "private" \emph{temporary domain}.

\subsubsection{Access Control}
TOS is an N-bit label added to each cache line's tag area. Each bit of TOS is bounded to a hardware thread (HT), a HT owns a cache line only if the corresponding bit is set. The ownership means the cache line can be accessed by the \emph{in-flight} operations of the HT. When a HT accesses the cache line it disowns, the \sys will emulate a latency equivalent to a cache miss and then set its corresponding TOS bit. On the other hand, if a HT evicts a cache line owned by other HTs, the \sys will simply reset the TOS bit instead of evicting the cache line. In summary, for each cache line in the temporary domain, \sys performs different actions according to the different situations:

\begin{packeditemize}
\item A thread can directly access a cache line it owns.
\item When a thread accesses a cache line that it does not own, \sys needs to check whether it is being owned by other threads or not: 
a) if it is \emph{not} owned by any other thread, which means it is invalid or unused, then the thread can install the requested data in it; 
b) if it is being owned by another thread, then a cache miss will be emulated by \sys to hide its true access latency. In either case, after the above process, the cache line will be marked as being owned by this thread.
\item When a cache line needs to be replaced or squashed in a thread's \emph{in-flight} operation, \sys marks the thread as no longer owning this cache line, and checks whether other threads own this cache line or not. If they do, \sys does nothing; otherwise, \sys replaces or squashes this cache line. 
\item When a cache line needs to be committed in a thread's \emph{in-flight} operation, this cache line is directly switched to the persistent domain.
\end{packeditemize}

\begin{figure}[!t]
	\centering
 	\includegraphics[width=\columnwidth]{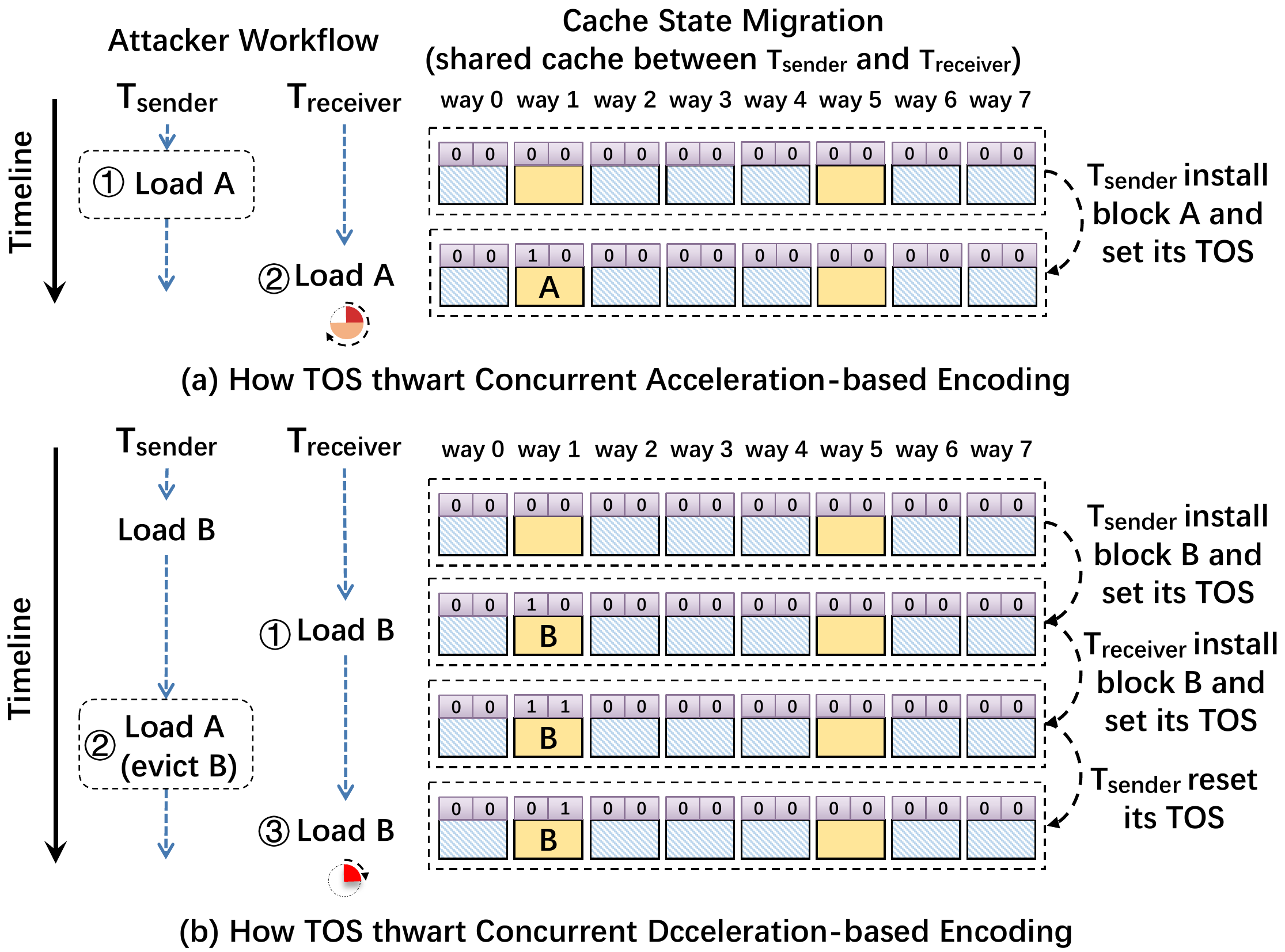}
 	\vspace{-0.8em}
	\caption{The workflow of the TOS scheme under two encoding attacks.}
	\vspace{-1.8em}
	\label{fig:TOS}
\end{figure}

\autoref{fig:TOS} shows how TOS mechanism can thwart the two encoding methods in the concurrent mode. For \emph{acceleration-based encoding}, a \emph{Load A} operation of $T_{sender}$ installs cache line A in the \emph{temporary} domain and set its TOS bit. Then, the $T_{receiver}$ also executes a \emph{Load A} operation. Because $T_{receiver}$ is not owning the cache line A, it will experience a latency emulated by the \sys, and regard it as a cache miss. 
For \emph{deceleration-based encoding} as shown in \autoref{fig:TOS}, the $T_{sender}$ first installs the cache line A in the temporary domain by a \emph{Load B} operation. Subsequently, the $T_{receiver}$ performs another \emph{Load B} operation. Once the $T_{receiver}$ completes, the TOS of the two threads are both set. And then the $T_{sender}$ performs a \emph{Load A} operation, which tries to replace the cache line B. At that time, \sys just resets the TOS of $T_{sender}$ instead of evicting the cache line, and suspends the \emph{Load A} request until it is committed or the $T_{receiver}$ resets the TOS also. Therefore, $T_{receiver}$ cannot detect the deceleration effect caused by $T_{sender}$.

\subsubsection{Implementation of TOS on different systems}
The TOS mechanism can also be easily implemented by adding labels in tag area and modifying the access controller of the cache.   
(1) For a hierarchical storage in a system that does not support SMT, the only component with concurrent issues is shared cache (i.e. LLC). The TOS can be efficiently implemented with the cooperation of cache-coherence directory and the T/P flag. When the T/P flag is P, the cache line can be accessed as usual because the TOS mechanism only targets speculative entries. When the T/P flag is T, the cache controller will process the accessing requests according to the above workflow using the ownership information recorded in the directory.
(2) For a hierarchy storage in a system that supports SMT, the private L1/L2 caches are also vulnerable. We need to add N bits to each cache line for the TOS, where N is the number of SMT threads. Each bit indicates the ownership of the corresponding thread. And for LLC, we also need to extend the directory and allow it to distinguish different SMT threads on each core. Therefore, for N SMT threads per core on an M-core system, the LLC needs to add $ (N-1) \cdot M $ bits for each entry.

\vspace{-0.5em}
\subsection{Other Enhanced Components}\label{sec:others}

In addition, \sys also blocks other potential attack surfaces from the auxiliary components in the cache system during the speculative execution, such as coherence states, cache management instructions and hardware prefetcher.

\begin{figure}[!t]
	\centering
 	\includegraphics[width=\columnwidth]{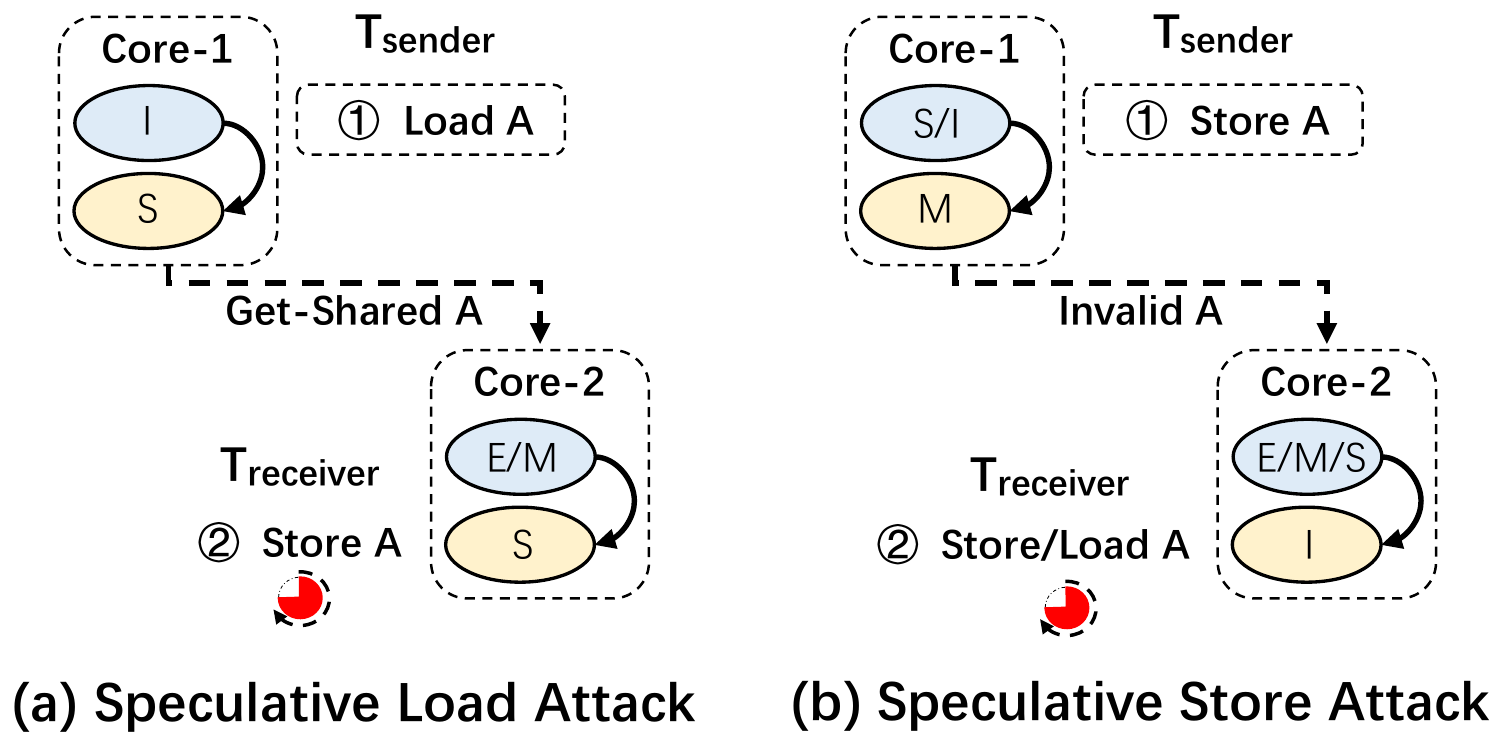}
	\caption{The coherence-state attacks.}
	\vspace{-1.8em}
	\label{fig:coherence}
\end{figure}

\bheading{Coherence States.}
The above-mentioned attacks use the cache layout as a medium for encoding. However, on a multi-core processor, cache coherence states may also be an attack surface. With the help of the \emph{Single-Writer-Multi-Reader} (SWMR) characteristic in the \emph{MESI} protocol, the attacker can complete \emph{deceleration-based encoding} attacks via the following two approaches~\cite{spectre_prime} as \autoref{fig:coherence} shows: (a) The $T_{sender}$ speculatively executes the load operation, which may causes the state of other cores’ copies changing from \emph{Exclusive} state to \emph{Shared} state, and then the $T_{receiver}$ will slow down if it exclusively accesses these copies again. (b) The $T_{sender}$ speculatively executes the store operation which may invalids other cores’ copies so that the $T_{receiver}$'s subsequent access will be missing. \sys leverage the scheme used in CleanupSpec~\cite{CleanupSpec} to address such a problem. It will delay the potentially malicious speculative state transactions, i.e. delaying the completion of the speculative load/store operations that may cause the above state changes in the MESI protocol. According to the evaluation of CleanupSpec~\cite{CleanupSpec} on some multi-thread workloads, the memory operations that may trigger the above two transactions are not frequent (about 2.4\%), so it will only introduce a small  performance overhead.

\bheading{Cache Management Instructions.}
In addition to speculative load/store instructions, an attacker can leverage some instructions for cache management during speculative execution to encode the secret, such as \emph{prefetch}, \emph{clflush} and \emph{INVD}. For these rare but potentially-vulnerable instructions, \sys stalls their execution, if they are issued during the speculative execution, and waits until they are committed to keep them from being used in TEAs.

\bheading{Hardware Prefetcher.}
\sys also supports \emph{hardware prefetching} because it is critical to the program performance. To ensure the security, \sys defers the speculative accesses to train the hardware prefetcher until they are committed. And the prefetched data will be directly installed into the \emph{persistent} domain.

\bheading{TLB and Paging Cache.}
TLB misses and Paging Cache misses may occur during the process of address translation for speculative memory accesses. Attackers can thus treat them as covert channels in TEAs. Considering the low frequency of TLB miss occurrence, \sys simply delays the execution of the \emph{in-flight} operation triggering TLB miss until it is committed. 

%% file: analysis.tex
\section{Security Analysis}\label{sec:analysis}

\begin{table}[!t]
  \centering
  \caption{Defense principles against various abstract attack scenarios. ``A'' and ``B'' represent thread A and  other threads that share the cache with thread A, respectively; ``t'' represents the temporary domain; ``p'' represents the persistent domain; ``evict'' means a thread evicts a cache line; ``install'' means a thread installs a cache line; ``access'' means a thread accesses a cache line. \emph{Assuming the secret is a 1-bit value, the expected measured result for an attacker can be either ``fast'' or ``slow''}. }
  \resizebox{\columnwidth}{!}{
  \setlength\tabcolsep{1.5pt}
    \begin{tabular}{cccccl}
    \toprule
    \textbf{Line}  & \textbf{Mode}  & \textbf{$S_{prepare}$} & \textbf{$S_{send}$} & \textbf{$S_{receive}$} & \textbf{Defense Principles} \\ \hline \midrule
    1     & \multirow{3}[6]{*}{\rot{Serialized}} & $A^{evict}_{t/p}$ & $A^{install}_{t}$ & $A_{t/p}$ (fast) & 
    \tabincell{l}{The cache line installed in $S_{send}$ \\ will be cleaned up before $S_{receive}$.}  \\ \cmidrule{3-6}   
    2     &       & $A^{install}_{p}$ & $A^{evict}_{t}$ & $A_{t/p}$ (slow) & 
    \tabincell{l}{The cache line in the persistent domain \\ cannot be evicted (replaced) in $S_{send}$.}  \\ \cmidrule{3-6} 
    3     &       & $A^{install}_{t}$ & $A^{evict}_{t}$ & $A_{t/p}$ (slow) & 
    \tabincell{l}{The cache line installed into the temporary \\ domain by an \emph{in-flight} operation is evicted \\ later. It will be reinstalled into the persistent \\ domain when that \emph{in-flight} operation \\ needs to be committed.} \\ \cmidrule{2-6}    
    4     & \multirow{5}[10]{*}{\rot{Concurrent}} & $B^{evict}_{t/p}$ & $A^{install}_{t}$ & $B_{t/p}$ (fast) &
    \tabincell{l}{When thread B firstly accesses the cache \\ line installed into the temporary domain \\ by thread A in $S_{receive}$, the cache miss \\ will be raised.} \\ \cmidrule{3-6}    
    5     &       & $B^{install}_{t}$ & $A^{evict}_{t}$ & $B_{t/p}$ (slow) & 
    \tabincell{l}{The cache line installed into the temporary \\ domain by thread B cannot be evicted by \\ thread A in $S_{send}$.} \\ \cmidrule{3-6}   
    6     &       & $B^{install}_{p}$ & $A^{evict}_{t}$ & $B_{t/p}$ (slow) & 
    \tabincell{l}{The cache line in the persistent domain \\ cannot be evicted by thread A in $S_{send}$.} \\
    \bottomrule
    \end{tabular}%
       }
    \vspace{-1.8em}
  \label{tab:principle}%
\end{table}%

After analyzing various existing TEAs, such as Spectre-PHT/BTB/RSB/STL~\cite{Spectre-PHT1,Spectre-PHT2,maisuradze2018ret2spec,koruyeh2018spectre,Spectre-STL}, Lazy FP~\cite{lazyfp}, MDS~\cite{mds1,mds2,mds3}, LVI~\cite{LVI} and CacheOut~\cite{CacheOut}, we abstract the transmission process of persistent convert channel as a three-stage model: $S_{prepare}$$\leadsto$$S_{send}$$\leadsto$$S_{receive}$. $S_{prepare}$ represents the stage of preparing the covert channel component. $S_{send}$ represents the stage of accessing secret through speculative execution and encoding the secret into the covert channel. $S_{receive}$ represents the stage of extracting and decoding the information from the covert channel. 

Based on the above model, \autoref{tab:principle} enumerates all possible actions of each stage (i.e. evict or install operation performs on any domain) and the expected states of $S_{receive}$ (i.e. slow or fast of execution time), and shows how \sys can block them. In \sys, we assume that the attacker knows all protection schemes and can manipulate any item in any domain during the $S_{prepare}$. However, because the speculative execution of $S_{send}$ is illegal and will eventually be squashed, the attacker can ONLY manipulate the data in temporary domain. For serialized mode attacks, the reliable timing method of $S_{receive}$ decide that it MUST begin after $S_{prepare}$ and $S_{send}$ complete (i.e. be committed in ROB). But for concurrent mode attacks, the order constraint of $S_{send}$ and $S_{prepare}$ can be relaxed. In the followings, we will elaborate the process in each attack mode and \sys's response via a representative example.

\bheading{Serialized-Mode Attacks.} 
Take the third attack (Line 3 in~\autoref{tab:principle}) as an example. In the $S_{prepare}$ stage, thread A installs a cache line into the temporary domain through an \emph{in-flight} operation. In the $S_{send}$ stage, thread A evicts this cache line via the replacement. In the $S_{receive}$ stage, thread A times the access to this cache line. With \sys, when the \emph{in-flight} operation in $S_{prepare}$ is committed, if the corresponding cache line is not present, \sys will reinstall it. So the slowdown of the access time cannot be measured in $S_{receive}$.

\bheading{Concurrent-Mode Attacks.} 
Take the sixth attack (Line 6 in the table) as an example. In the $S_{prepare}$ stage, thread B installs a cache line. In the $S_{send}$ stage, thread A performs an \emph{in-flight} operation to evict this cache line. In the $S_{receive}$ stage, thread B times the access to this cache line. With \sys, the cache line will not be evicted in $S_{send}$ because thread B owns this cache line. The access to this cache line by thread B in $S_{receive}$ will hit the cache, and the slowdown of the access time cannot be measured.

As mentioned before, besides normal data, attackers can also exploit some metadata as the covert channel, such as replacement trace bits or coherence state. Since \sys prevents the unsafe speculative modification of such metadata like other works, and carefully extends the cache without introducing new attack surface. \sys can also be used to defend against the metadata attacks, such as Speculative Interference~\cite{Speculativeinterference}.

%% file: eval.tex
\begin{table}[!h]
  \centering
  \caption{Parameters of simulated micro-architecture.}
  \resizebox{\columnwidth}{!}{
    \begin{tabular}{ll}
    \toprule
    \textbf{Parameter} & \textbf{Value} \\
    \hline
    \midrule
    Core  & 8-issue, out-of-order, 2Ghz \\
    \midrule
    Pipeline & \tabincell{l}{64-entry IQ, 192-entry ROB, 32-entry LQ, \\ 32-entry SQ, 256 Int / 256 FP registers}\\
    \midrule
        BPU   & Tournament branch predictor, 4096 BTB \\
    \midrule
        Private L1-I Cache & \tabincell{l}{32KB, 64B line, 4-way, 1 cycle RT latency, 4 MSHRs} \\
    \midrule
        Private L1-D Cache & \tabincell{l}{64KB, 64B line, 8-way, 1 cycle RT latency, 4 MSHRs} \\
    \midrule
        Shared L2 Cache & \tabincell{l}{2MB bank, 64B line, 16-way, 8 cycles RT local latency, \\16 cycles RT remote latency, 16 MSHRs} \\
    \midrule
        Coherence Protocol & inclusive, Directory-based MESI protocol \\
    \midrule
        Network & 4×2 mesh, 128b link width, 1 cycle latency per hop \\
    \midrule
        DRAM  & RT latency: 50 ns after L2 \\
    \bottomrule
    \end{tabular}%
    }
    \vspace{-1.5em}
  \label{tab:setup}%
\end{table}%

\section{Evaluation}\label{sec:eval}

\subsection{Experimental Setup}\label{sec:setup}

We implemented and simulated \sys based on an out-of-order (O3) processor and Ruby cache system using Gem5 \cite{Gem5}. The parameter settings of each component are shown in \autoref{tab:setup}, which are consistent with the settings in other studies. We evaluated the performance using 27 benchmarks from SPEC CPU 2006 and 12 benchmarks from PARSEC-3.0. The reason some benchmarks such as \emph{dealII} and \emph{tonto} are not included is because they cannot be simulated correctly by the version \emph{fe187de9bd} of Gem5 we used w/o \sys. For SPEC benchmarks, we use the \emph{ref} input set and skip the first 10 billion instructions in the fast-forward mode, and then perform cycle-level simulation on the next 1 billion instructions. For PARSEC benchmarks, under the full system with 8 cores, we use the \emph{simmedium} input set and simulate the instructions in the region of interest (ROI) at the cycle level. 

\subsection{Defense Against Spectre PoC Code}\label{sec:defense_effect}
\begin{figure}[!h]
	\centering
	\includegraphics[width=\columnwidth]{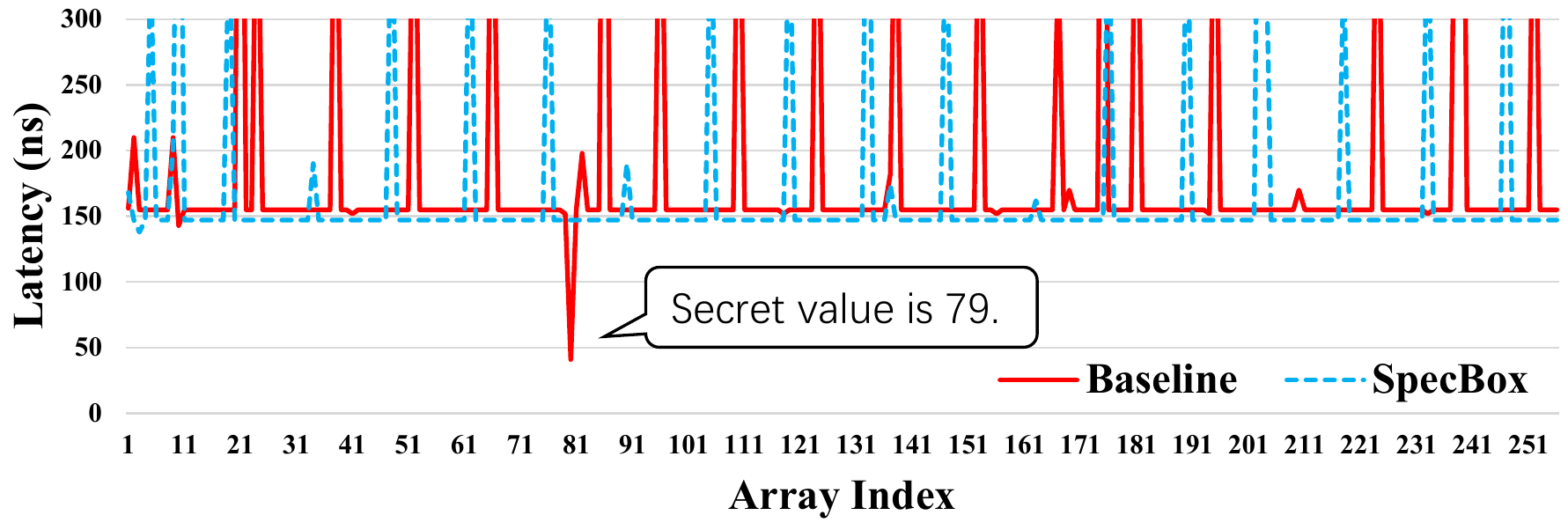}
	\caption{Access latency of the auxiliary array using \sys.}
	\vspace{-1.5em}
	\label{fig:attack}
\end{figure}

To evaluate the mitigation effectiveness of \sys, we chose the Spectre-PHT attack~\cite{Spectre-PHT1} as shown in \autoref{list:poc} that uses data cache as the covert channel. In the attack, we used a mis-predicted overflow access to obtain the secret whose value is 79, and then use the secret to index a 256-item auxiliary array. Finally, we measured the access latency of each array element 100 times. 
The access time is shown in \autoref{fig:attack}. We can see that when an access hits, the latency is below 50 cycles. In contrast, the latency exceeds 150 cycles if it misses. On the baseline processor without any protection, the attacker can clearly distinguish the access time difference of item 79 from other items in the array. However, using \sys, the item is evicted from the data cache when the transient instruction is squashed. Therefore the attacker can no longer distinguish the difference in access time between item 79 and others.

\subsection{Determine Domain Capacity}\label{sec:domain_capacity}

For \sys, the capacity ratio of the two domains is crucial to the performance because it determines the maximum number of the speculative and non-speculative cache lines. 

\begin{figure}[!h]
	\centering
	\includegraphics[width=\columnwidth]{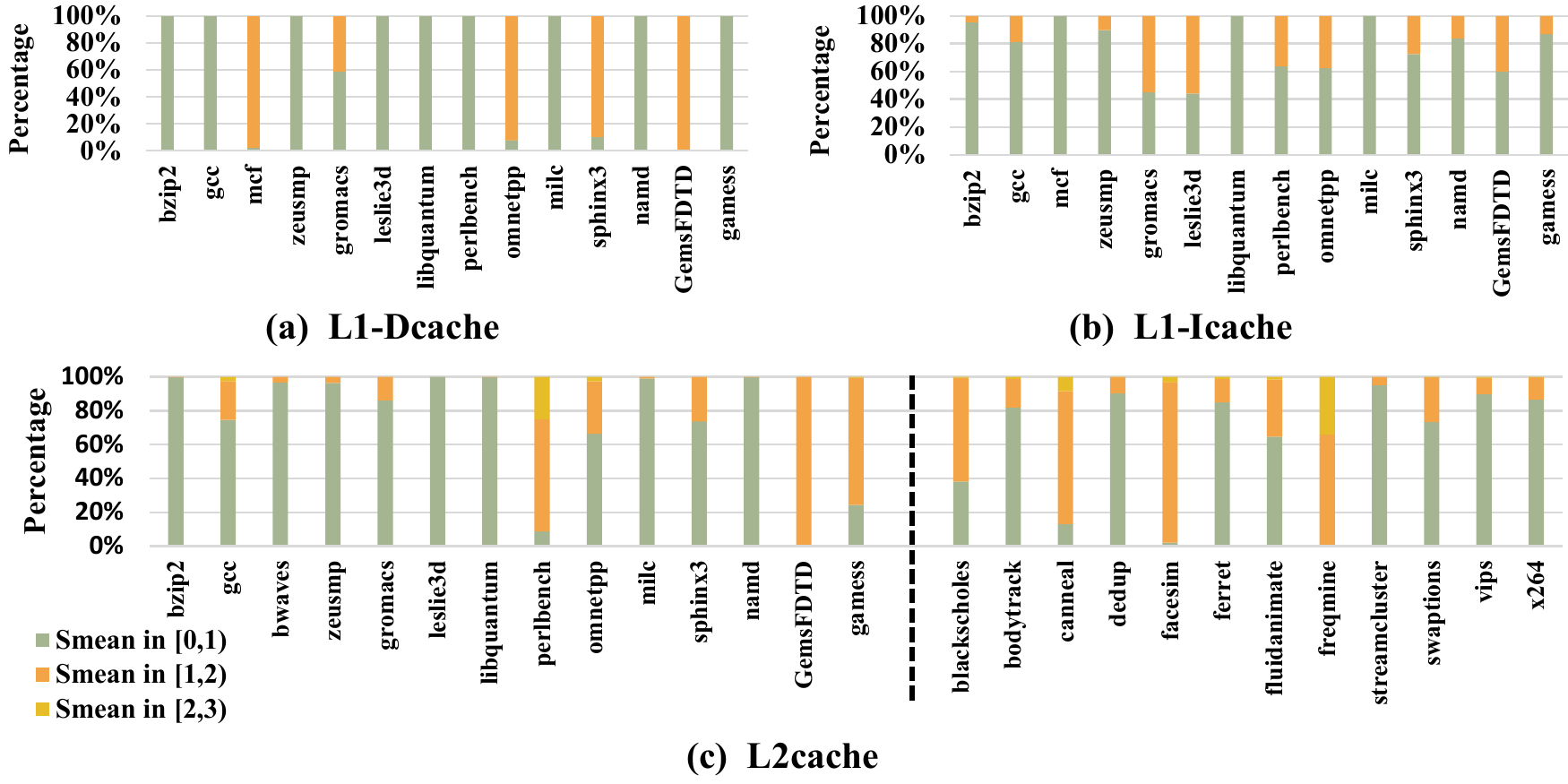}
	\vspace{-1.5em}
	\caption{We sampled a number of speculative cache lines in each cache set per Kilo instructions, and then calculate the mean of them as the representative value for each cache set (denoted as $S_{mean}$). The figure shows the percentage of the number of the cache sets with different $S_{mean}$. We found no cache set's $S_{mean}$ is larger then 3. We selected some benchmarks with distinguished distributions as examples, and the remaining benchmarks are similar to these examples.}
	\label{fig:speculative_number}
\end{figure}

Two factors need to be considered when determining the domain capacity. The first is that the temporary domain should be sufficiently large to contain the data installed by the \emph{in-flight} operations in any speculative window. These data will be frequently referenced within the current window and a period of time after being committed. \autoref{fig:speculative_number} shows the number of speculative cache lines in each cache sets sampled in the original cache system (i.e. w/o partitioning). 
We can see that, reserving 2 ways in L1-DCache and L1-ICache and 3 ways in the unified L2-Cache shared by multi-cores, are sufficient for most single- and multi-threaded programs. 

\begin{figure*}[!t]
	\centering
	\includegraphics[width=\textwidth]{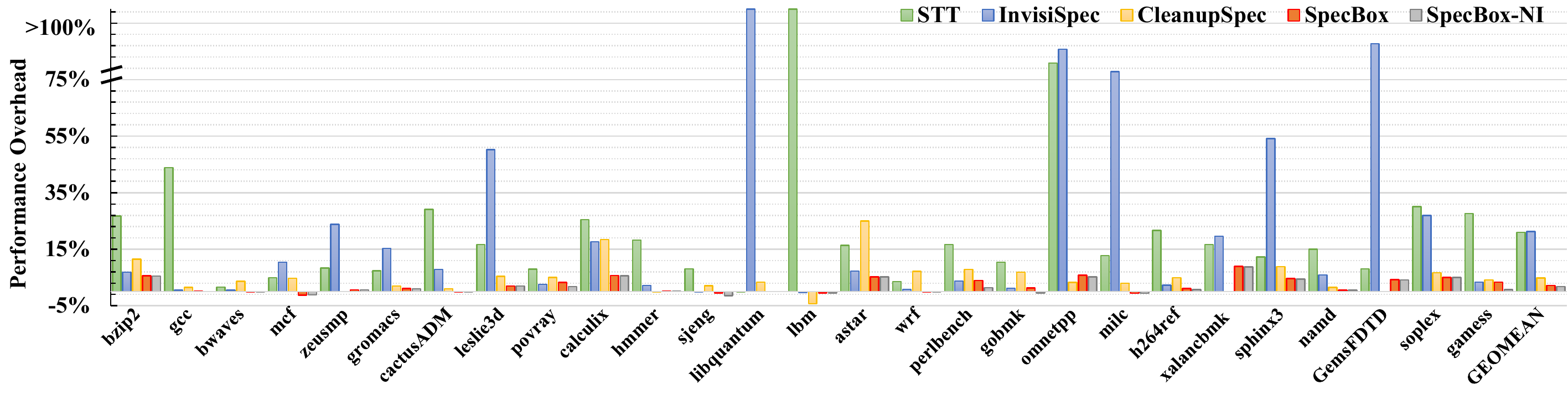}
	\vspace{-1.8em}
	\caption{The performance overhead of \sys, \sys with no-icache protected (denoted as \sys-NI), CLeanupSpec~\cite{CleanupSpec}, InvisiSpec~\cite{InvisiSpec} and STT~\cite{STT} on SPEC CPU 2006 benchmarks. For CleanupSpec, three benchmarks (i.e. \emph{zeusmp, xalancbmk, and GemsFDTD}) hang and failed in our simulation.}
	\vspace{-1.5em}
	\label{fig:performance_spec}
\end{figure*}

\begin{figure}[!h]
	\centering
	\includegraphics[width=\columnwidth]{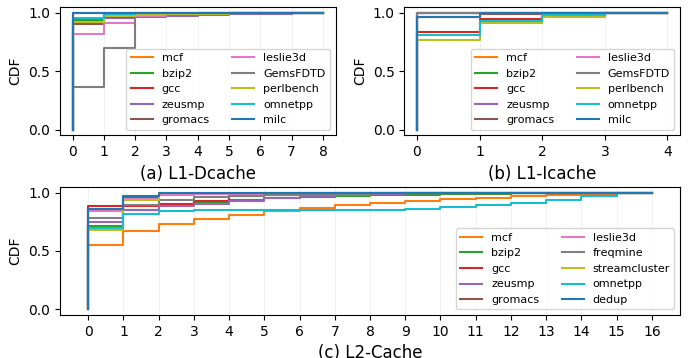}
	\vspace{-1em}
	\caption{We counted the distribution of hits in all cache sets and calculated the arithmetic mean as the representative distribution for each benchmark. The horizontal axis in the figure represents the cache lines in one cache set, which are arranged in LRU order from left to right, and the vertical axis is the cumulative distribution of hit counts. We selected some benchmarks with distinguished distributions as examples, and the remaining benchmarks are similar to these examples}
	\vspace{-1.5em}
	\label{fig:hit_distribution}
\end{figure}

The second is that the temporary domain should not be too large to deprive the capacity of the persistent domain. It is because the persistent domain contains all of the non-speculative and committed data in the cache. Its capacity is more sensitive to the programs that have more access patterns with long reuse distances, such as \emph{mcf} and \emph{GemsFDTD} in SPEC. Therefore, we counted the number of accesses to different ways in each cache set, and calculate the cumulative distribution in original (non-partitioning) cache system. The access order is decided by the replacement policy (e.g. LRU in \sys). As the \autoref{fig:hit_distribution} shows, for most programs, over 90\% of accesses hit the first three ways in L1-DCache and L2-Cache, and the first way in L1-ICache. It thus will not impact the cache utilization much if we take away the last 2 or 3 ways for the temporary domain. This observation is consistent with other approaches that adopt way-partitioning schemes~\cite{Catalyst}.

\vspace{-1em}

\subsection{Performance Overhead}\label{sec:performance}

We compared \sys with other open-sourced hardware-based defenses for TEAs. The first is STT\cite{STT}. It represents the scheme that delays the instructions that are dependent on the transient instructions. The second is InvisiSpec\cite{InvisiSpec}. It represents the scheme that cleans up the side effect of transient instructions. The third is CleanSpec~\cite{CleanupSpec}. It represents the scheme that rollbacks the cache layout when speculation fails. In order to have a fair comparison with \sys, for both InvisiSpec and STT, we choose the \emph{futuristic model} that can resist all Meltdown-type and Spectre-type attacks; 
for CleanupSpec, we choose the rollback strategy for L1-DCache and L2-Cache. Because InvisiSpec and CleanupSpec are currently not used in the instruction cache, we implemented a version of \sys only on data cache and not on instruction cache (i.e. \sys-NI) to compare with other schemes.

\begin{figure}[!t]
	\centering
	\includegraphics[width=\columnwidth]{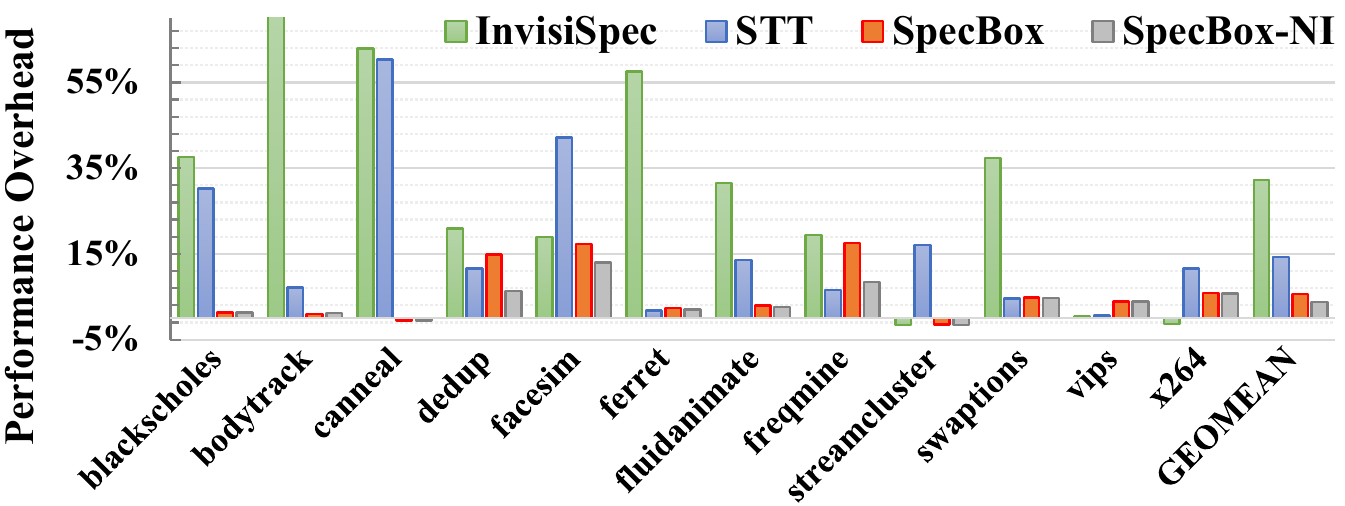}
	\vspace{-1.5em}
	\caption{The performance overhead of InvisiSpec, STT, SpecBox and SpecBox-NI  on PARSEC 3.0 benchmarks. CleanupSpec currently does not support simulating multi-threaded applications in Gem5.}
	\vspace{-1.8em}
	\label{fig:performance_parsec}
\end{figure}

In \autoref{fig:performance_spec} and \autoref{fig:performance_parsec}, we can see that the overall performance overheads of \sys for SPEC and Parsec are 2.17\% and 5.61\%, respectively, which are much lower than 20.97\% and 32.27\% with STT, 21.29\% and 11.62\% with InvisiSpec, and also better than CleanupSpec (4.77\% for SPEC). The largest overheads of \sys are 8.96\% for \emph{GemsFDTD} in SPEC and 14.96\% for \emph{debup} in PARSEC, which are also better than InvisiSpec and STT. When we disable the protection on the instruction cache (i.e. L1-ICache and the corresponding regions in L2-Cache), the performance overheads of \sys are reduced to 1.76\% and 3.85\%, respectively.

\bheading{\sys versus STT.}
The main overhead of STT comes from pipeline stalls caused by the delayed execution. Therefore, for benchmarks with more dependent instructions that can cause pipeline stalls such as \emph{lbm, ommnetpp} and \emph{gcc} in SPEC, and \emph{swaptions, ferret} and \emph{bodytrack} in PARSEC, the performance of STT is inferior to InvisiSpec, CleanupSpec and \sys. On the other hand, STT has the advantage that, if an instruction such as the load instruction whose dependent instructions have become safe, STT can simply lift the protection and allow it to modify the states for replacement and coherence. It has much less overhead for those programs with less dependence but with a large amount of cache accesses, such as \emph{sphinix} and \emph{GemsFDTD} in SPEC and \emph{streamcluster} in PARSEC.

\bheading{\sys versus InvisiSpec.}
The main cost of InvisiSpec comes from reload operations when a speculative instruction is committed. The main intention of the reload operation is to prevent subsequent replacement of the committed data in Speculative Buffer (SB) from being exploited by attackers. Another intention is that the SB is private to each core and, hence, will not receive invalidation requests from other cores, which may lead to the violation of the memory consistency model. It needs to reload the data from the cache hierarchy when the \emph{in-flight} load instruction is committed, and validate the current value with the used value during the speculative execution. If the validation fails, it must squash and re-execute the instructions following that load instruction. From the \autoref{fig:performance_spec}, we can see that for memory-intensive applications, such as \emph{leslie3d, GemsFDTD and bip2}, the performance differences between InvisiSpec and other schemes are particularly distinct. The situation is more serious in some PARSEC benchmarks such as \emph{canneal} and \emph{facesim}. It is because more invalidation requests are generated in those benchmarks due to cache coherence transactions in a multi-core system, hence, the validation failures and re-execution will occur more frequently.

\bheading{\sys versus CleanupSpec.}
CleanupSpec addresses the problem in InvisiSpec by allowing the replacement of a  speculative instruction and restoring the layout if it is squashed. This scheme works well for most of the programs, such as \emph{libquantum} and \emph{omnetpp} in SPEC benchmarks. But for some programs with higher misprediction rates, such as \emph{astar}, \emph{bzip2} and \emph{gobmk}, it will incur a certain amount of performance overhead (10\%-20\%), which is larger than that in InvisiSpec and \sys. And for some cases with low misprediction rates but need to process subsequent squashed instructions, like \emph{calculix} and \emph{sphnix3}, the stall from rollback operations will also degrade the performance.


\subsection{Main Cost Analysis}

\subsubsection {Squashes due to wrong-path execution}
The essence of caching is to exploit the space and temporary locality in programs to overcome the memory wall. The unchanged size of cache line guarantee the spatial locality not be broken by \sys. The first change brought by \sys is eliminating the temporary locality of the data installed by the wrong path execution, which is also used in InvisiSpec and CleanupSpec. Previous studies have shown that wrong-path execution has the following two side effects~\cite{wrong_path}.

\begin{figure}[!t]
	\centering
	\includegraphics[width=\columnwidth]{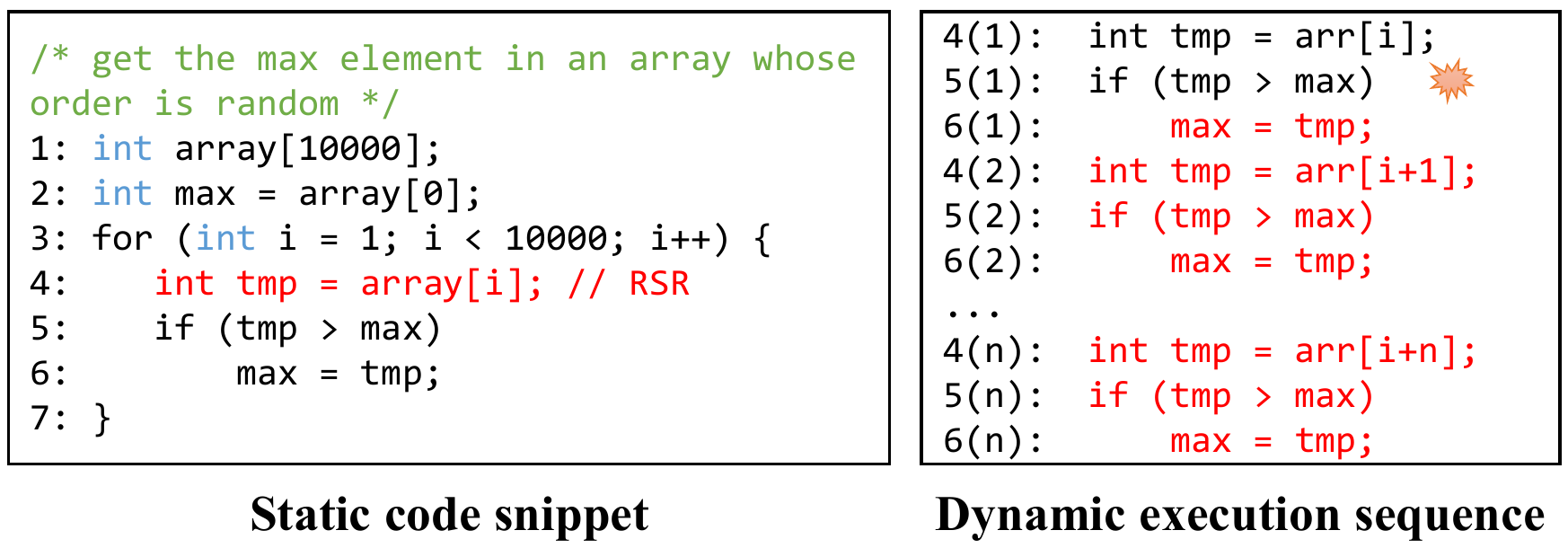}
	\vspace{-1.5em}
	\caption{A simplified code snippet with a RSR access pattern.}
    \vspace{-1.5em}
	\label{fig:rsr}
\end{figure}

\begin{figure}[!h]
	\centering
	\includegraphics[width=\columnwidth]{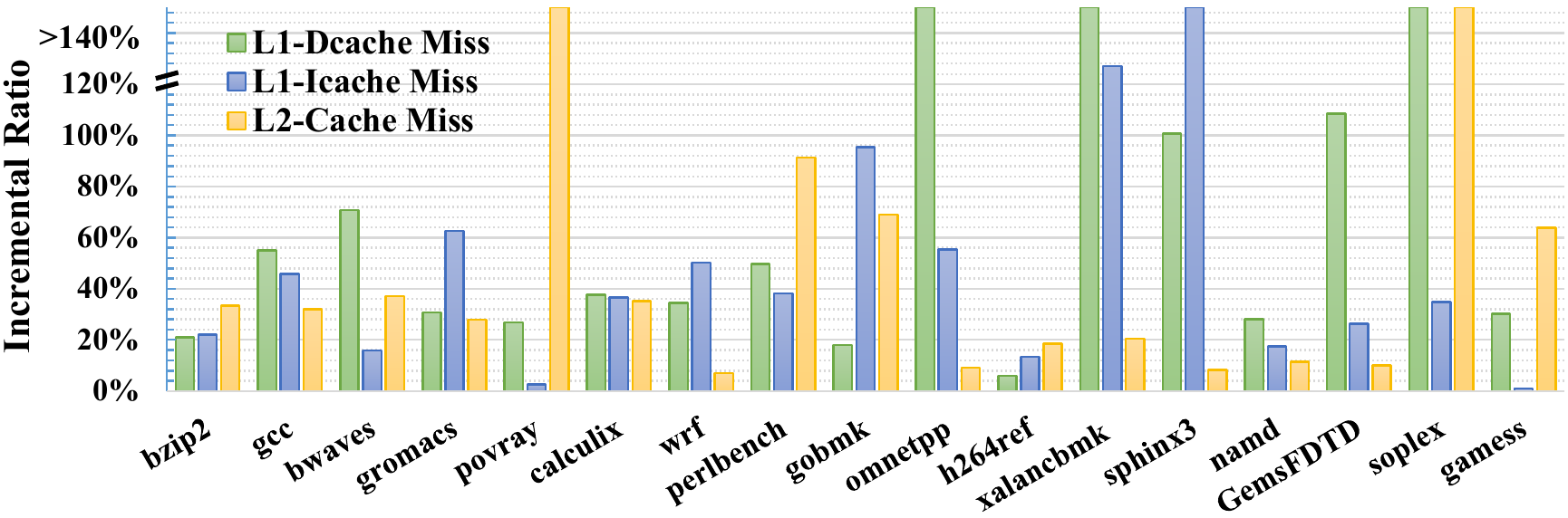}
	\vspace{-1.8em}
	\caption{The incremental ratio of cache misses from the committed memory operations in the original cache system and those in \sys w/o domain capacity limitation.}
	\vspace{-1em}
	\label{fig:squash_cache_miss}
\end{figure}

\bheading{Prefetching Effect.} 
The data loaded during wrong-path reference will be re-referenced by the future correct-path execution. \autoref{fig:rsr} gives an simplified example of this scenario. The code in the example tries to scan an array and find its maximum element. In an out-of-order execution, the statements in Lines 4-6 from several iterations of the loop will be executed in the same speculative window. However, the array elements are un-ordered, the branch prediction on Line 5 will fail frequently, so the subsequent 4(2), 4(3), …, 4(n) instructions will be squashed. In \sys, we call such a sequence \emph{Reference-Squash-Rereference} (RSR). \autoref{fig:squash_cache_miss} shows the incremental ratio of cache misses from the committed memory instructions in each cache level  without limiting the domain capacity, and only perform cleanup operations for the \emph{temporary} domain. From the results, We can see that RSR access patterns exist in most workloads. Especially, when they occur frequently in L1-DCache, it will cause a certain degree of performance overhead as shown in \emph{xalancbmk} and \emph{soplex} in SPEC.

\bheading{Pollution Effect.}
The data loaded by wrong-path accesses will never be re-referenced but will replace other useful data. Furthermore, these wrong-path accesses will trigger the hardware prefetcher, which may pollute the cache indirectly. This situation rarely occurs in general out-of-order processors, and usually only occurs in more aggressive optimizations, such as run-ahead execution~\cite{run_ahead}. However, in \sys, the domain isolation scheme and secure prefetcher scheme alleviate the pollution effect caused by the wrong-path accesses.

\subsubsection{TOS for multi-core systems}
\begin{figure}[!t]
	\centering
	\includegraphics[width=\columnwidth]{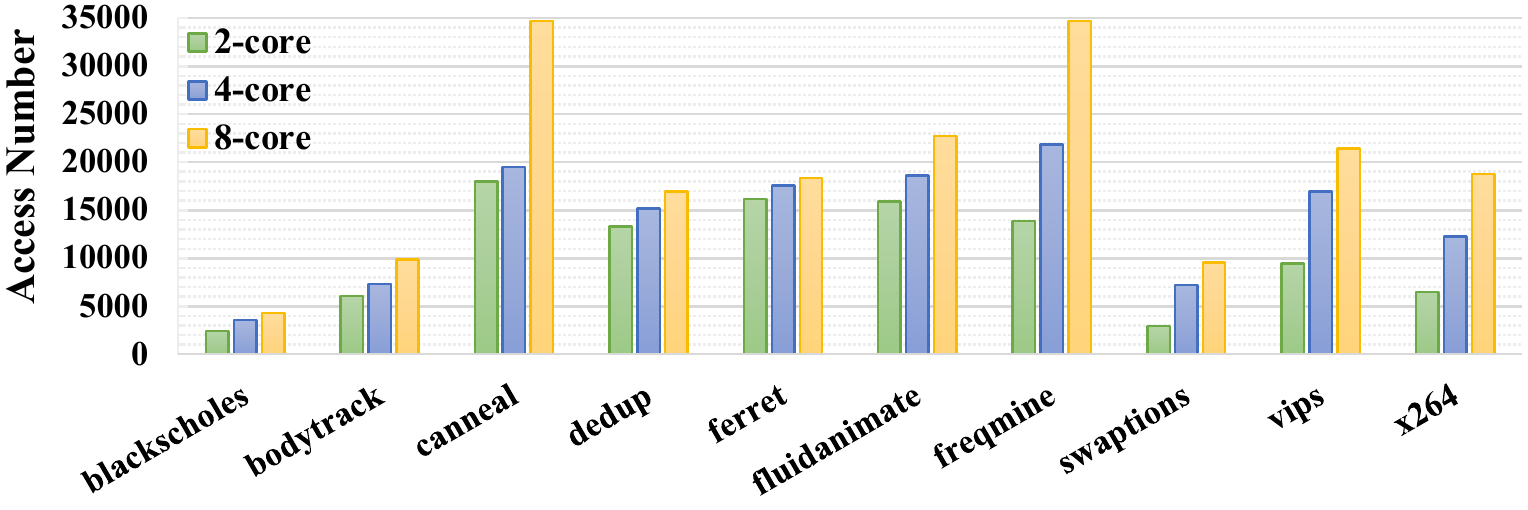}
	\vspace{-1.8em}
	\caption{The number of simultaneous speculative accesses to L2-Cache in PARSEC on different number cores. Some cores access the cache line for the first time.}
	\vspace{-1.5em}
	\label{fig:simultaneous_access}
\end{figure}

In multi-core systems with \sys, one important source of overheads comes from the emulated latency needed to support TOS scheme when different cores perform speculative accesses to a shared cache line simultaneously. The latency is not added to all simultaneous accesses but only to the first access from a core. \autoref{fig:simultaneous_access} shows the number of such accesses to L2-Cache in systems with different number of cores. We can see that as the number of cores increases, the number of such accesses gradually increases. And in cases such as \emph{canneal} and \emph{freqmine}, TOS scheme does have some impact there.

In addition, as the number of cores increases, due to the limited temporary domain capacity, the competition among multiple cores for the same cache set will increase. This is because TOS does not allow the cache line in the speculative state to be unilaterally evicted by a single owner. \autoref{fig:scale} shows the performance of most PARSEC benchmarks gradually decreases as the number of cores increases (note: \emph{facesim} and \emph{dedup} crashed or hang, thus not included). But for some benchmarks, such as \emph{bodytrack, ferret} and \emph{vips}, when the number of cores exceeds 8, there is a significant drop in performance.

\subsection{Hardware Cost and Power Consumption}

In order to implement the \emph{domain partitioning mechanism}, \sys adds 1 bit to each cache line in the tag array. To implement the \emph{thread ownership semaphores (TOS)} for L1-ICache, L1-DCache on a core with two physical SMT threads, \sys adds additional 2 bits for each cache line in the tag array. For the shared L2 cache, the \sys adds 2, 4, 8 bits, respectively, to implement TOS on 2-core, 4-core, and 8-core systems. We used CACTI-6.5 \cite{CACTI} to evaluate the hardware cost and the power consumption for the additional storage requirements. The results are shown in \autoref{tab:power}. As we can see, the \sys introduces modest hardware and power consumption overhead.


\begin{table}[!h]
  \centering
  \caption{The hardware and power overhead of \sys.}
  \vspace{-0.5em}
  \resizebox{\columnwidth}{!}{
    \begin{tabular}{lllll}
    \toprule
    \multicolumn{1}{c}{Metric} & \multicolumn{1}{c}{L1-ICache} & \multicolumn{1}{c}{L1-DCache} & \multicolumn{1}{c}{L2-Cache (2/4/8-core)}  \\
    \midrule
    Area & 0.18\%  & 0.38\%  & 0.61\% / 0.61\% / 0.92\% \\
    \midrule
    Access time & 0.15\% & 0.01\% & 3.03\% / 3.03\% / 3.31\% \\
    \midrule
    Dynamic read energy & 0.77\% & 0.63\% & 0.05\% / 0.05\% / 0.10\% \\
    \midrule
    Dynamic write energy & 4.72\% & 2.21\% & 0.01\% / 0.01\% / 0.14\% \\
    \midrule
    Leakage power & 1.52\% & 0.93\% & 2.09\% / 2.09\% / 2.47\% \\
    \bottomrule
    \end{tabular}%
  }
  \vspace{-1.5em}
  \label{tab:power}%
\end{table}

%% file: discuss.tex
\section{Discussion}\label{sec:discuss}




\bheading{Domain Capacity Configuration.}
As shown in our evaluation (see \ref{sec:domain_capacity}), a pre-determined capacity ratio can satisfy the requirement of most programs with today's cache sizes and associativity. However, for some performance-sensitive applications, dynamic adjustment of the domain capacity can provide more flexibility. \sys uses a set of privilege registers \emph{domain\_cap} for capacity re-configuration at each level cache. The registers can only be controlled by special serialized instructions to avoid being controlled by illegal speculative execution. When the capacity of the temporary domain becomes zero, it means that the system gives up all security protection. Thus, the cache will treat all access as non-speculative accesses, and the notifier will suspend squash/commit requests to the cache.

\begin{figure}[!t]
	\centering
	\includegraphics[width=\columnwidth]{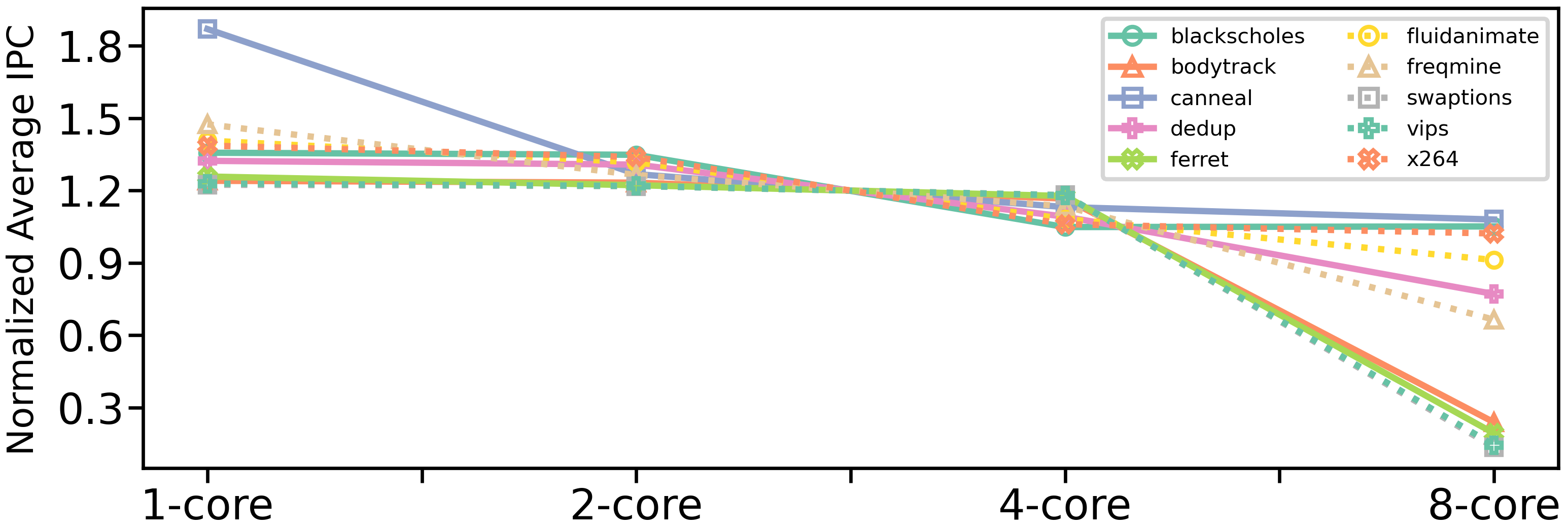}
	\vspace{-1em}
	\caption{The normalized IPC of the multi-core system with \sys.}
	\vspace{-1em}
	\label{fig:scale}
\end{figure}

\bheading{Extending on other components.}
In addition to the cache system, \sys can also be extended to other storage components that can be exploited as persistent covert channel in TEAs. To reduce power consumption, Intel processors will turn off the high bit of an execution unit if the AVX2 instruction is not executed for a long time, and turn it back on when it is executed. The turn-on operation will take a lot of time, resulting in a difference in execution time, which can be exploited by attackers~\cite{schwarz2019netspectre}. In this case, we can also add TOS to AVX2 units. When a thread turns on the high bit of the AVX2 unit in an \emph{in-flight} operation, the corresponding bit in the TOS is set to 1. At this time, if another thread also needs to perform the AVX2 instruction, and its corresponding TOS bit is 0, it will wait for a latency that is equivalent to the duration of turning it on. When a thread has not used the AVX2 instruction for a while, the CPU will first set the corresponding bit in the TOS to 0, and wait until all of the N bits in the TOS are all 0, then it can turn off the high bit. 

\bheading{Defense against TSA.}
To defend against serial-mode attacks, \sys prevents the receiver from detecting any side effect after the sender exits its speculative window. However, assume that the receiver and sender are active in the \emph{same} speculative window and the length of the speculative window can be measured, the attacker may still be able to gather side effects of the sender execution. SpectreRewind~\cite{SpectreRewind} has shown that such attacks are feasible in practice, which leverages contention on an execution unit as a covert channel. However, its variant that leverages contention on the cache (which is the focus of \sys) has yet to be shown feasible. 
Such potential vulnerability also exists in other existing defenses against cache covert channels. SafeSpec~\cite{SafeSpec} discusses this vulnerability (called \emph{Transient Speculative Attack} (TSA)), and proposes two possible solutions.
One solution is to change the speculative storage to fully associative, and increase its capacity to avoid contention.
The other solution is to partition the speculative storage to accommodate sender and receiver branches separately. Both solutions will incur some hardware overhead.
The second solution can be extended in \sys, which will be our future work.

%% file: relwk.tex
\section{Related Work}\label{sec:relwk}

In addition to the \emph{delaying} and \emph{invisible speculation} approaches mentioned in \autoref{sec:defenses}, there are other defenses achieved through other software or hardware means.

\bheading{Preventing Speculative Execution on Sensitive Code.}
The most direct way to prevent TEAs is to insert Fence~\cite{fence} or SSBB~\cite{SSBB} instructions in the security-sensitive code. Moreover, Spectre-type attacks can be prevented by preventing the control-flow prediction unit from being trained by attackers. IBRS/STIBP/IBPB~\cite{IBRS,STIBP,IBPB} and Retpoline~\cite{Retpoline} prevent branch prediction unit of different permissions and different threads from interfering each other. CSF~\cite{CSF} modifies the decode stage in pipeline to automatically inject fence instructions before branch instructions. These defenses has a simple defense principle but may cause serious performance degradation compared with \sys.

\bheading{Preventing Illegal Speculative Accesses.}
Another type of mitigation is to prevent unauthorized instructions from obtaining the secret’s value during speculative execution. KPTI~\cite{KPTI} separates the page table entries and TLB entries in user-space from those in kernel-space. Chrome and Webkit browsers~\cite{SiteIsolation} prevent cross-site transient access through index masking and pointer poisoning. OISA~\cite{OISA} ensures that the accesses to sensitive data must use special instructions from a customized instruction subset, and these instructions cannot be executed out of order. ConTExT~\cite{ConTExT} marks the protected memory pages and registers to prevent their data from being obtained in the out-of-order execution state.  Comparing with these defenses, \sys can protect all the secret data from Spectre attacks instead of the special data assigned by software. 

\bheading{Partition for Software Domains.}
Partitioning based on processes or other software domains have been used to block existing side-channel attacks. Some prior work~\cite{Chameleon, STEALTHMEM, h1, h6} uses cache set partitioning to enhance security via page coloring on physical-page allocation, which may lead to possible high memory overhead. Catalyst~\cite{Catalyst} adopts Intel’s CAT~\cite{CAT}, a hardware-supported way partitioning scheme on the last-level cache, to protect the Xen hypervisor with a low memory overhead. SecDCP~\cite{secdcp} dynamically adjusts domain sizes according to the number of incurred cache misses to improve the performance of CAT partitioning. DAWG~\cite{DAWG} improves cache way partitioning to enhance the isolation for hits, misses and metadata across the domains. It also provides an optimized software solution for secure domain time-multiplexing, cache synonyms avoidance and efficient cross-domain data transfer. However, compare with \sys, all these defenses require programmers to annotate the secrets for protection and cannot thwart the transient-execution attacks (TEAs) within the domain.

\bheading{Randomization and Other Approaches.}
Randomization and other noise-injection schemes are another effective approach to defend against side-channel attacks. For example, CEASER~\cite{CEASER} and RPcache~\cite{RPcache} can randomize the mapping of the addresses, cache sets or TLB sets, which prevent attackers from preparing the data layouts in the target cache set. RFillCache~\cite{rfill_cache} selects random victims during cache-line replacement to hide the side effects of cache replacement policies. Secure-TLB~\cite{Secure-TLB} adopts a similar randomization replacement strategy to defeat TLBleed attacks~\cite{TLBleed}. FTM~\cite{FTM} targets cross-core Flush+Reload attacks by delaying the first access to the last-level cache. All these defenses can only be applied to a limited set of covert channels. However \sys is a more general defense approach for all types of persistent covert channel in TEA.


%% file: conclusion.tex
\vspace{-0.5em}
\section{Conclusion}\label{sec:conclusion}
This paper presented a \emph{label-based transparent speculation} scheme, called \sys, to defend against transient execution attacks. 
It partitions each level cache into a \emph{temporary} and a \emph{persistent} domain, by attaching a 1-bit label to each item and isolate the side effect of \emph{in-flight} and \emph{committed} operations. An \emph{in-flight} operation can only affect the temporary domain, and the affected items will be switched to the persistent domain when the operation needs to be committed. To avoid the change of the temporary domain being observed by other synchronous executing threads, \sys introduces \emph{thread ownership semaphores}, which dynamically marks the thread ownership of each (shared) item in the temporary domain and emulates a thread-private storage. Analysis and extensive experiments have shown that \sys is not only secure, but also practical and efficient.

%% file: ack.tex
\section*{Acknowledgements}\label{sec:ack}
This research was supported by the National Natural Science Foundation of China (NSFC) under grant 61902374 and U1736208. Pen-Chung Yew is supported by the NSF under
the grant CNS-1514444. Zhe Wang is the corresponding author (wangzhe12@ict.ac.cn). Any opinions, findings, conclusions or recommendations expressed in this material are
those of the authors and do not necessarily reflect the views of NSF.